\documentclass[sigconf, nonacm]{acmart}

\usepackage{multirow}
\usepackage{adjustbox}
\usepackage{booktabs}
\usepackage{amsfonts}
\usepackage{arydshln}
\usepackage{xspace}

\usepackage{pifont}
\newcommand{\cmark}{\ding{51}}

\newcommand{\uniform}{\textsc{Uniform Calibration}\xspace}
\newcommand{\hard}{\textsc{CLS-Hard Calibration}\xspace}
\newcommand{\soft}{\textsc{CLS-Soft Calibration}\xspace}

\AtBeginDocument{%
  }

\setcopyright{acmlicensed}
\copyrightyear{2018}
\acmYear{2018}
\acmDOI{XXXXXXX.XXXXXXX}
\acmConference[Preprint]{Make sure to enter the correct
  conference title from your rights confirmation email}{Preprint}{Zurich, CH}
\acmISBN{978-1-4503-XXXX-X/2018/06}

\begin{document}

\title{Attention Calibration for Position-Fair Dense Retrieval}

\author{%
  Andrianos Michail, Elias Schuhmacher, Juri Opitz, Simon Clematide, Rico Sennrich
}

\affiliation{%
  \institution{University of Zurich}
  \city{Zurich}
  \country{Switzerland}
}

\email{%
  andrianos.michail,simon.clematide, rico.sennrich@uzh.ch
}

\renewcommand{\shortauthors}{Michail et al.}



\received{20 February 2007}
\received[revised]{12 March 2009}
\received[accepted]{5 June 2009}

\begin{abstract}
Dense retrieval compresses a passage into a single vector, but this compression
is positionally skewed: early content dominates the embedding, and retrieval
degrades when the relevant span appears later. Prior work proposed an inference-time method that
counteracts this skew by equalizing the pooling token's attention across passage segments. 
However, (i) it redistributes attention at a fixed strength, (ii) it
forces the pooling token's attention to itself to a fixed basket-level mass despite substantial variation across layers and architectures, and
(iii) its effect on retrieval has not been evaluated. 
We introduce a strength coefficient that interpolates between uncalibrated and fully equalized attention, together with an efficient implementation that reduces peak calibration memory overhead from 5–7\,GiB to under 1\,MiB.
Across three embedding models and two pooling
schemes, moderate calibration provides a better retrieval trade-off than full equalization. We introduce a
 variant that preserves the pooling token's self-attention mass
and redistributes only the remaining mass. On a position-aware retrieval benchmark
spanning 10 languages and 31 domains, a configuration selected on English FineWeb-PosQ and transferred without tuning reduces position
sensitivity in all 16 evaluated length-quartile, model, and retrieval-setting
combinations, by up to 43\% relative, while improving nDCG@10 by up to 4.8\% relative and
leaving general retrieval effectiveness on NanoBEIR essentially unchanged. Calibration runs at indexing time, adding
no query-time latency. We release our code at \href{https://github.com/impresso/fair-sentence-transformers}{github.com/impresso/fair-sentence-transformers}
\end{abstract}

\maketitle

\begin{figure}[t] 
\centering
\includegraphics[width=1.0\linewidth, trim=25 2 18 2, clip]{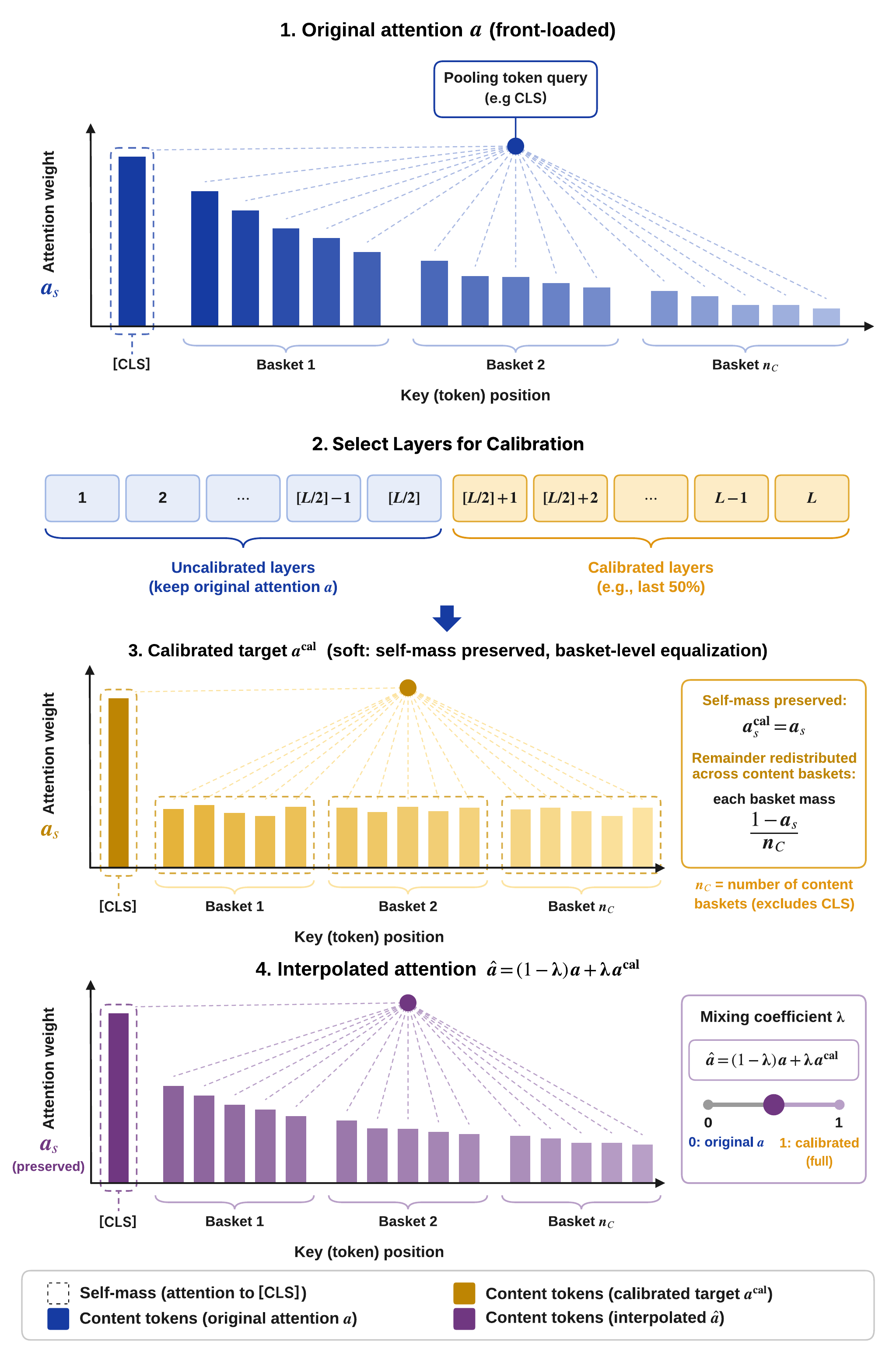} 
\caption{Attention calibration with strength coefficient $\lambda$. In the
calibrated layers $\mathfrak{L}^C$, the pooling token's front-loaded attention
$\alpha$ is replaced by $\alpha^{\mathrm{cal}}$, which preserves the self-mass
$a_s$ and equalizes the remainder across content baskets (\soft). $\lambda \in [0,1]$ interpolates between them.}
\Description{A four-panel schematic. The first panel shows the pooling token's
original, front-loaded attention as a bar chart over key positions: a node for
the pooling token's query fans out to all keys, a tall separate bar at the left
marks its attention to itself, and the remaining bars are tallest in the first
basket and fall off steadily across the later baskets. The second panel is a
row of boxes representing the transformer layers, with the first half marked as
uncalibrated and the last half as calibrated. The third panel shows the
calibrated target: the self-attention bar is unchanged while the content bars
are close to equal height across baskets, each basket receiving an equal share
of the remaining mass. The fourth panel shows the interpolated result, lying
between the first and third, with a slider for the strength coefficient lambda
running from the original distribution at zero to the fully calibrated one at
one. A legend distinguishes the self-attention bar from the original,
calibrated and interpolated content bars.}
\label{fig:sample_frontpage}
\end{figure}

\section{Introduction}
Dense retrieval models are central to modern information retrieval systems,
mapping queries and documents into a shared embedding space for efficient
semantic matching. A passage, however long, is compressed into a single vector,
so its content competes for limited representational space
--- and that competition is not neutral. \citet{coelho-etal-2024-dwell} show that dense embedding models exhibit
\emph{positional bias}: content early in a passage is disproportionately
reflected in the embedding. \citet{zeng-etal-2025-empirical} confirm the
consequence for retrieval, where effectiveness degrades when relevant
information appears later in a passage.

\citet{schuhmacher2026informationrepresentationfairnesslongdocument} trace this
imbalance to the pooling token's attention, which is front-loaded toward early
key positions, and propose an inference-time calibration that partitions the key
sequence into contiguous baskets and equalizes attention mass across them. The
appeal is practical: no retraining or architectural changes, and calibration operates at encoding time.

That method, however, was designed and evaluated for a rep\-re\-sen\-ta\-tion-level
objective --- how evenly document positions are reflected in the embedding --- rather than retrieval effectiveness.  We find two limitations. Full equalization can overcorrect,
improving late-position retrieval at a larger cost to early positions, and the
method offers no way to control its strength. It also overwrites the pooling token's
attention to itself with a fixed constant, although this quantity varies
substantially across layers and architectures.

We address both limitations. A \emph{strength
coefficient} $\lambda \in [0,1]$ interpolates between the uncalibrated and the fully equalized attention distributions, allowing direct control over calibration strength (Figure~\ref{fig:sample_frontpage}). We also introduce a \emph{self-mass-aware} redistribution method, \soft, which preserves the pooling token’s attention to its own position and redistributes only the remaining mass. Finally, calibration requires the full post-softmax attention matrix, which adds 5–7\,GiB at peak for a batch of 2{,}048-token passages in the original implementation. Our targeted implementation reduces this overhead to under 1\,MiB, making systematic evaluation practical.

We use FineWeb-PosQ~\citep{zeng-etal-2025-empirical} to characterize how calibration strength, calibrated layer set
$\mathfrak{L}^C$, basket size $\mathfrak{B}$, and redistribution variant affect position sensitivity and retrieval effectiveness, and to select a single configuration. Moderate calibration provides a better retrieval trade-off than full equalization, and the selected configuration requires no per-model tuning across the evaluated \texttt{<s>}-pooled and last-token-pooled architectures. Transferred
without modification to the Position-Aware IR Benchmark (PosIR)~\citep{zeng2026posir}, it reduces
position sensitivity in every evaluated combination of length quartile, model,
and retrieval setting across 10 languages and 31 domains, while largely preserving or
improving retrieval effectiveness.
NanoBEIR performance remains essentially unchanged. A chunk-and-average baseline achieves comparable sensitivity reductions only at the cost of several nDCG points.

We make the following contributions:
\begin{itemize}
\item \textbf{Calibration with tunable strength:}
We extend the method of
\citet{schuhmacher2026informationrepresentationfairnesslongdocument} with a
strength coefficient $\lambda \in [0,1]$ that interpolates between the uncalibrated and fully equalized attention distributions. Moderate calibration gives a better retrieval trade-off than full equalization used in prior work. An efficient implementation that
operates only on the modified attention rows reduces peak memory
overhead from 5--7\,GiB to under 1 MiB, making systematic parameter exploration
tractable for long documents.

\item \textbf{Self-mass-aware redistribution:}
The pooling token's attention to itself varies substantially across layers and architectures, making a fixed $1/n_{\mathrm{baskets}}$ allocation potentially distortive. We therefore treat this mass separately from positional redistribution: \soft preserves it and redistributes only the remainder, while  \hard fixes it at a model-independent value. We adopt \soft as our default and show it applies to both \texttt{<s>}-pooled and last-token-pooled models.

\item \textbf{Systematic study of calibration parameters:}
We analyze how $\lambda$, $\mathfrak{L}^C$, $\mathfrak{B}$, and the
redistribution variant jointly affect positional
sensitivity and retrieval effectiveness, identifying settings in which calibration improves
or degrades retrieval.

\item \textbf{Shared configuration and large-scale evaluation:}
We select a shared configuration on FineWeb-PosQ  and evaluate it on
PosIR across 10 languages, 31 domains, and two embedding models. It
consistently reduces position sensitivity while largely preserving or improving
retrieval effectiveness in both multilingual and cross-lingual
(x$\rightarrow$en) settings. Calibration operates only at indexing time, adding no
query-time latency.
\end{itemize}

\section{Related Work}

\paragraph{Position Bias in Dense Retrieval.}
\citet{zeng-etal-2025-empirical,zeng2026posir} introduced position-aware evaluation sets to study how the location of relevant information within a passage affects dense retrieval effectiveness. The first dataset is called \textit{FineWeb-PosQ.} The authors sampled 13{,}902 passages of 500--1{,}024 words from FineWeb-edu, divided each into beginning, middle, and end segments, and generated section-specific questions with GPT-4o-mini, resulting in 25{,}775 questions. Models exhibit reduced retrieval performance for the middle and end segments. The more recent  \textit{Position-Aware IR Benchmark} (PosIR) extends this evaluation to multilingual and cross-lingual retrieval. It comprises 310 datasets spanning 10 languages and 31 domains, with relevance tied to precise reference spans. Its central methodological feature is a length-controlled bucketing strategy that groups queries by the length of the positive document (Q1: 512, Q2: 1024, Q3: 1536, Q4: 2048 tokens) and analyzes positional effects within each bucket, isolating position bias from length-induced degradation. Results across ten state-of-the-art embedding models show that position bias is pervasive and intensifies with document length. Both benchmarks document position bias, but neither evaluates a mitigation. We use FineWeb-PosQ to select a configuration and PosIR to test whether that choice survives a change of language, domain, and document length.

\paragraph{Attention Calibration.}
\citet{schuhmacher2026informationrepresentationfairnesslongdocument} analyze how information from different document positions contributes to the final embedding representation.
Using a permutation-based evaluation framework, they show that multilingual encoder-based embedding models exhibit systematic positional imbalance: early segments are over-represented in the document embedding, while later segments are progressively less reflected.
They attribute this effect to front-loaded attention distributions of the pooling token, whose query assigns disproportionate attention mass to early key tokens across transformer layers.
Their evaluation, however, focuses on representation-level positional balance rather than retrieval effectiveness. We examine whether calibration improves retrieval and whether its effects extend to models that pool on the final token rather than the first.

\section{Method}


\subsection{The Starting Point: Attention Calibration}

Most modern embedding models designate one position as the \emph{pooling token}: a prepended \texttt{<s>} in encoder-based models, or the final token \texttt{</s>} in last-token-pooled models. Its representation is read out as the embedding of the whole sequence, so its attention distribution affects how information from different positions contributes to the embedding. We also use CLS as shorthand for the pooling token, and the variant names in Section~\ref{sec:selfattn} follow this convention.

The original inference-time attention calibration proposed by \citet{schuhmacher2026informationrepresentationfairnesslongdocument} redistributes the pooling token's attention more evenly across document positions. Specifically, the method partitions the key sequence into contiguous baskets of size $\mathfrak{B}$ and enforces equal total attention mass across baskets while preserving relative attention weights within each basket. The pooling token is isolated in its own basket to avoid redistributing its attention mass across content tokens. For a token at position $k$ assigned to basket $b(k)$, the calibrated attention weight is:
\[
a^{\mathrm{cal}}_k = \frac{a_k}{\sum_{j \in b(k)} a_j} \cdot \frac{1}{n_{\mathrm{baskets}}},
\]
where $a_k$ is the original post-softmax attention weight and $n_{\mathrm{baskets}}$ is the number of baskets. Calibration is applied only to the pooling token's query row in selected transformer layers and requires no retraining.

Two choices are fixed in the original formulation: redistribution is always complete, and the pooling token's own basket always receives $1/n_{\mathrm{baskets}}$. Sections~\ref{sec:strength} and~\ref{sec:selfattn} relax each in turn.

\begin{figure}[t] 
\centering
\includegraphics[width=0.99\linewidth, trim=21 65 40 26, clip]{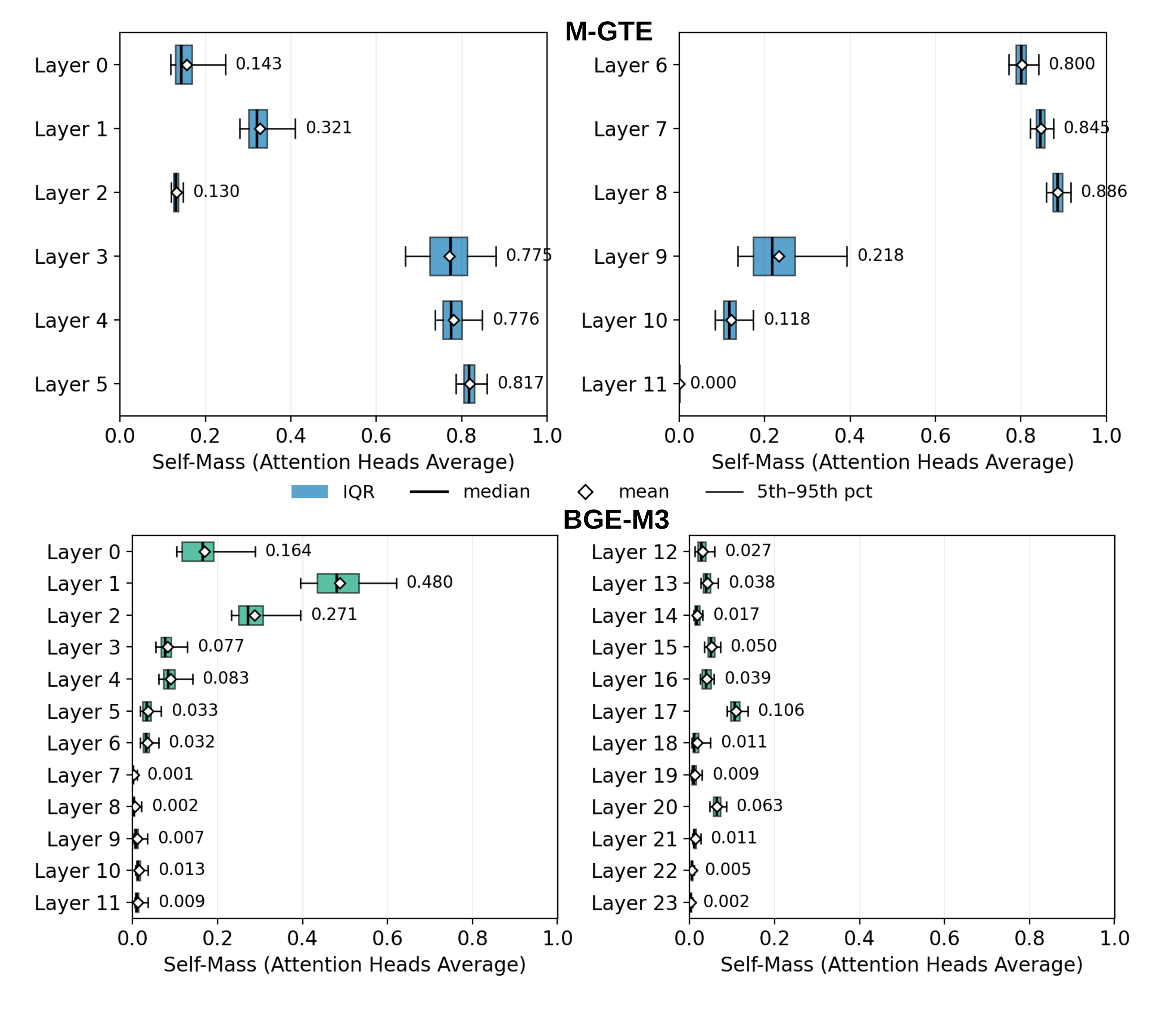} 
\caption{Pooling-token self-mass $a_s$ across layers.}
\Description{Box plots of the pooling token's self-mass, one row per layer, on
a scale from 0 to 1. For gte-multilingual-base the values are low in the first
three layers, then rise above 0.75 and stay high through the middle and later
layers. For bge-m3-dense the values peak below 0.5 in the early layers and drop
close to zero from roughly layer 6 onwards.}
\label{fig:cls-selfattn}
\end{figure}

\subsection{Calibration Strength Coefficient}
\label{sec:strength}
First, we generalize the attention calibration method of \citet{schuhmacher2026informationrepresentationfairnesslongdocument} with a strength coefficient $\lambda \in [0, 1]$ that linearly interpolates between the uncalibrated and calibrated attention distributions of the pooling token. The resulting attention weight for key position $k$ is:
\[
\hat{a}_k = \lambda \cdot a^{\mathrm{cal}}_k + (1 - \lambda) \cdot a_k,
\]
where $\lambda = 0$ recovers the original model and $\lambda = 1$ applies full calibration as in the original formulation. The coefficient allows us to vary the strength of positional redistribution and examine the trade-off between position sensitivity and retrieval effectiveness. Implementation details are provided in Appendix~\ref{app:implementation}. We calibrate passage/document encodings only; query encodings are left unchanged.

We refer to this variant as \textbf{\uniform}, since the fully calibrated distribution assigns equal attention mass across baskets.

\begin{figure*}[t]
\centering
\includegraphics[width=0.9\textwidth,trim=8.5mm 0 12mm 10,clip]{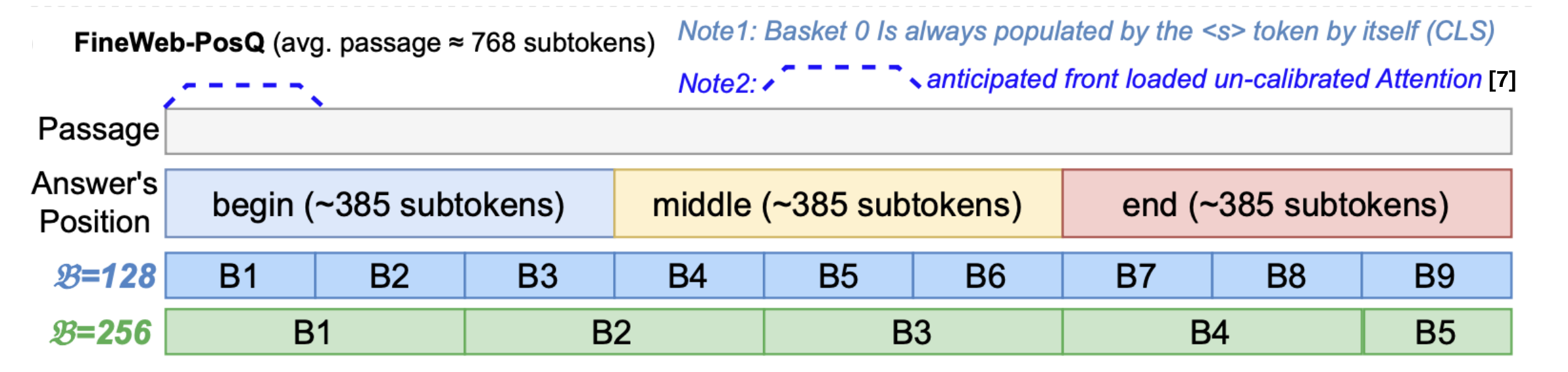} 
\caption{Evaluation segments and attention calibration baskets for FineWeb-PosQ. Baskets partition the token sequence independently of evaluation groupings; $\mathfrak{B}$ determines the size of baskets and the granularity of redistribution.}
\Description{A schematic of one passage drawn as a horizontal bar. Below it,
one row divides the passage into three equal evaluation segments labelled
begin, middle and end. Two further rows show the calibration baskets over the
same span: nine equal baskets at a basket size of 128 tokens, and five at a
basket size of 256. The basket boundaries do not line up with the segment
boundaries.}
\label{fig:schematic}
\end{figure*}

\subsection{Self-Mass-Aware Redistribution}
\label{sec:selfattn}

The original formulation isolates the pooling token in its own basket, assigning it $1/n_{\mathrm{baskets}}$ of the attention mass regardless of its original share. Figure~\ref{fig:cls-selfattn} shows that in an uncalibrated forward pass this quantity fluctuates strongly across layers and differs between models: in
gte-multilingual-base the pooling token directs 0.80--0.89 of its attention mass
to itself in the middle layers, while in bge-m3-dense it stays below 0.11
throughout the second half of the network. 
A fixed budget of
$1/n_{\mathrm{baskets}}$ --- 0.16 at six baskets, 0.08 at twelve --- is close to
neither and, in the first case, removes most of the mass  allocated
to the pooling token itself. This adjustment is not motivated by positional balance, and the
discrepancy grows with the number of baskets, i.e.\ with document length.

Concentration of attention on particular positions has been documented in transformer models.
\citet{xiao2024efficient} describe it as an \emph{attention sink} --- the self-mass is the sink mass of the calibrated row: softmax
forces each row to sum to one, so surplus attention accumulates on a
consistently positioned token, in encoder models as well as decoders. The mass
carries little content, but disturbing it rescales everything else. Reducing the
pooling token's share from $a_s$ to $1/n_{\mathrm{baskets}}$ multiplies every
content weight by $(1 - 1/n_{\mathrm{baskets}})/(1 - a_s)$ --- a factor of 5.6 at
$a_s = 0.85$ and six baskets --- pushing the content distribution well outside
the model's own range, for reasons unrelated to position.

We therefore treat the pooling tokens's self-mass  separately from 
positional redistribution. Let $a_s$ denote the weight the pooling token's query
assigns to its own key position --- the diagonal entry of the calibrated row --- which we call its \emph{self-mass}, and let $\mathfrak{B}^C$ be the set of content baskets.

\textbf{\soft:} This variant of calibration retains $a_s$ and equalizes
only the remaining mass across content baskets:
\[
a^{\mathrm{cal}}_k = \frac{a_k}{\sum_{j \in b(k)} a_j} \cdot
\frac{1 - a_s}{|\mathfrak{B}^C|}, \qquad k \neq s .
\]
The redistribution budget thus follows the model's self-mass at each
layer  and adapts to 
different self-mass profiles across architectures.

\textbf{\hard:} The two variants above determine $a_s$ differently: under \uniform it depends on the basket count, and hence on document length, while \soft preserves
the model's original allocation. \hard instead fixes $a_s$ independently of both, assigning the pooling token a fixed weight $w$ and equalizing the remainder across content baskets:
\[
a^{\mathrm{cal}}_s = w, \qquad
a^{\mathrm{cal}}_k = \frac{a_k}{\sum_{j \in b(k)} a_j} \cdot
\frac{1 - w}{|\mathfrak{B}^C|}, \qquad k \neq s .
\]
We set $w = 0.3$ throughout as a moderate fixed level,
independent of both the model and the number of baskets. Because the models we
examine sit on opposite sides of it, the same constant acts in opposite
directions, lowering the self-mass in one and raising it in
the other. \hard therefore tests whether a fixed level suffices, or
whether tracking the model's own allocation is what matters.

All variants use $\lambda$ to interpolate between the original
distribution and their respective calibrated distributions.

\begin{table*}[t]
\centering
\small
\renewcommand{\arraystretch}{0.95}
\setlength{\tabcolsep}{4pt}
\setlength{\dashlinedash}{0.5pt}
\setlength{\dashlinegap}{1.5pt}
\resizebox{\textwidth}{!}{%
\renewcommand{\arraystretch}{0.941}
\begin{tabular}{l ccccc @{\hspace{12pt}} ccccc @{\hspace{12pt}} ccccc}
\toprule
\textbf{Indexing Method}
& \multicolumn{5}{c}{\textbf{\uniform}}
& \multicolumn{5}{c}{\textbf{\soft}}
& \multicolumn{5}{c}{\textbf{\hard}} \\
\cmidrule(lr){2-6} \cmidrule(lr){7-11} \cmidrule(lr){12-16}
& begin & mid & end & HM & PSI $\downarrow$
& begin & mid & end & HM & PSI $\downarrow$
& begin & mid & end & HM & PSI $\downarrow$ \\
\midrule

\multicolumn{16}{l}{\textbf{gte-multilingual-base (340M)}} \\
\hspace{0.8em} No Calibration & 87.16 & 78.01 & 72.23 & 78.66 & 0.171 & \multicolumn{10}{c}{---} \\
\hspace{0.8em} Segment Embed Average & 76.88 & 78.83 & 74.65 & 76.75 & 0.053 & \multicolumn{10}{c}{---} \\
\cmidrule[0.2pt](lr){1-6}
\hspace{0.8em} last layer, $\lambda=0.5$, $\mathfrak{B}=128$ & 86.59 & 79.15 & 74.00 & 79.58 & 0.145 & 86.58 & 79.36 & 74.20 & 79.73 & 0.143 & 86.47 & 78.76 & 73.57 & 79.25 & 0.149 \\
\hspace{0.8em} last layer, $\lambda=0.5$, $\mathfrak{B}=256$ & 86.45 & 78.94 & 73.90 & 79.43 & 0.145 & 86.46 & 79.12 & 74.48 & 79.72 & 0.139 & 86.41 & 78.67 & 73.66 & 79.24 & 0.148 \\
\addlinespace[1.5pt]
\hspace{0.8em} last layer, $\lambda=1.0$, $\mathfrak{B}=128$ & 85.70 & 79.97 & 74.55 & 79.81 & 0.130 & 85.65 & 80.38 & 75.17 & 80.17 & 0.122 & 85.43 & 78.76 & 73.27 & 78.84 & 0.142 \\
\hspace{0.8em} last layer, $\lambda=1.0$, $\mathfrak{B}=256$ & 85.58 & 79.26 & 74.29 & 79.44 & 0.132 & 85.49 & 80.05 & 75.54 & 80.15 & 0.116 & 85.33 & 78.52 & 73.38 & 78.78 & 0.140 \\
\addlinespace[1.5pt]
\hspace{0.8em} 50\% layers, $\lambda=0.5$, $\mathfrak{B}=128$ & 84.96 & 79.57 & 74.56 & 79.47 & 0.122 & 85.91 & 80.18 & 75.29 & 80.23 & 0.124 & 85.45 & 79.49 & 74.36 & 79.51 & 0.130 \\
\hspace{0.8em} 50\% layers, $\lambda=0.5$, $\mathfrak{B}=256$ & 84.86 & 79.21 & 74.63 & 79.35 & 0.121 & 85.71 & 79.78 & 75.60 & 80.15 & 0.118 & 85.13 & 79.12 & 74.57 & 79.37 & 0.124 \\
\addlinespace[1.5pt]
\hspace{0.8em} 50\% layers, $\lambda=1.0$, $\mathfrak{B}=128$ & 79.39 & 79.56 & 74.18 & 77.63 & 0.068 & 82.87 & 81.86 & 76.90 & 80.46 & 0.072 & 79.47 & 79.90 & 74.49 & 77.87 & 0.068 \\
\hspace{0.8em} 50\% layers, $\lambda=1.0$, $\mathfrak{B}=256$ & 79.22 & 79.21 & 75.12 & 77.80 & 0.052 & 82.68 & 81.22 & 77.73 & 80.49 & 0.060 & 79.31 & 79.32 & 75.42 & 77.97 & 0.049 \\
\midrule[0.3pt]
\multicolumn{16}{l}{\textbf{bge-m3-dense (560M)}} \\
\hspace{0.8em} No Calibration & 88.77 & 78.39 & 71.88 & 79.09 & 0.190 & \multicolumn{10}{c}{---} \\
\hspace{0.8em} Segment Embed Average & 73.72 & 77.62 & 72.77 & 74.65 & 0.062 & \multicolumn{10}{c}{---} \\
\cmidrule[0.2pt](lr){1-6}
\hspace{0.8em} last layer, $\lambda=0.5$, $\mathfrak{B}=128$ & 88.11 & 79.29 & 73.15 & 79.72 & 0.170 & 88.14 & 79.38 & 73.33 & 79.83 & 0.168 & 88.28 & 78.98 & 72.78 & 79.51 & 0.176 \\
\hspace{0.8em} last layer, $\lambda=0.5$, $\mathfrak{B}=256$ & 88.04 & 79.08 & 73.14 & 79.63 & 0.169 & 88.05 & 79.31 & 73.46 & 79.83 & 0.166 & 88.27 & 78.94 & 72.88 & 79.54 & 0.174 \\
\addlinespace[1.5pt]
\hspace{0.8em} last layer, $\lambda=1.0$, $\mathfrak{B}=128$ & 87.21 & 79.85 & 74.14 & 80.05 & 0.150 & 87.23 & 80.03 & 74.49 & 80.25 & 0.146 & 87.58 & 79.42 & 73.35 & 79.70 & 0.162 \\
\hspace{0.8em} last layer, $\lambda=1.0$, $\mathfrak{B}=256$ & 87.10 & 79.57 & 74.20 & 79.95 & 0.148 & 87.07 & 79.95 & 74.94 & 80.35 & 0.139 & 87.44 & 79.34 & 73.66 & 79.75 & 0.158 \\
\addlinespace[1.5pt]
\hspace{0.8em} 50\% layers, $\lambda=0.5$, $\mathfrak{B}=128$ & 86.36 & 80.27 & 75.17 & 80.34 & 0.130 & 86.62 & 80.55 & 75.47 & 80.62 & 0.129 & 86.68 & 79.92 & 74.46 & 80.04 & 0.141 \\
\hspace{0.8em} 50\% layers, $\lambda=0.5$, $\mathfrak{B}=256$ & 85.79 & 79.72 & 75.59 & 80.15 & 0.119 & 86.22 & 80.11 & 76.19 & 80.63 & 0.116 & 86.34 & 79.58 & 75.19 & 80.11 & 0.129 \\
\addlinespace[1.5pt]
\hspace{0.8em} 50\% layers, $\lambda=1.0$, $\mathfrak{B}=128$ & 75.41 & 79.61 & 76.72 & 77.21 & 0.053 & 77.30 & 80.13 & 77.18 & 78.18 & 0.037 & 65.25 & 72.04 & 69.25 & 68.73 & 0.094 \\
\hspace{0.8em} 50\% layers, $\lambda=1.0$, $\mathfrak{B}=256$ & 75.02 & 78.74 & 77.57 & 77.08 & 0.047 & 78.06 & 79.87 & 78.70 & 78.87 & 0.023 & 71.59 & 76.27 & 75.12 & 74.27 & 0.061 \\

\midrule[0.3pt]
\multicolumn{16}{l}{\textbf{Qwen3-Embedding-0.6B}} \\
\hspace{0.8em} No Calibration & 88.54 & 78.83 & 65.61 & 76.49 & 0.259 & \multicolumn{10}{c}{---} \\
\hspace{0.8em} Segment Embed Average & 77.67 & 79.22 & 77.05 & 77.97 & 0.027 & \multicolumn{10}{c}{---} \\
\cmidrule[0.2pt](lr){1-6}
\hspace{0.8em} last layer, $\lambda=0.5$, $\mathfrak{B}=128$ & 88.53 & 78.87 & 65.81 & 76.59 & 0.257 & 88.40 & 78.56 & 65.08 & 76.13 & 0.264 & 88.48 & 78.87 & 65.70 & 76.53 & 0.257 \\
\hspace{0.8em} last layer, $\lambda=0.5$, $\mathfrak{B}=256$ & 88.47 & 78.86 & 65.78 & 76.56 & 0.256 & 88.41 & 78.62 & 65.05 & 76.14 & 0.264 & 88.45 & 78.84 & 65.72 & 76.52 & 0.257 \\
\addlinespace[1.5pt]
\hspace{0.8em} last layer, $\lambda=1.0$, $\mathfrak{B}=128$ & 87.92 & 78.61 & 65.59 & 76.26 & 0.254 & 87.89 & 78.00 & 64.19 & 75.42 & 0.270 & 87.88 & 78.51 & 65.40 & 76.13 & 0.256 \\
\hspace{0.8em} last layer, $\lambda=1.0$, $\mathfrak{B}=256$ & 87.92 & 78.58 & 65.61 & 76.26 & 0.254 & 87.90 & 77.98 & 63.94 & 75.30 & 0.273 & 87.90 & 78.52 & 65.45 & 76.16 & 0.255 \\
\addlinespace[1.5pt]
\hspace{0.8em} 50\% layers, $\lambda=0.5$, $\mathfrak{B}=128$ & 83.46 & 81.15 & 75.63 & 79.94 & 0.094 & 83.13 & 81.18 & 75.14 & 79.67 & 0.096 & 82.99 & 80.28 & 73.35 & 78.66 & 0.116 \\
\hspace{0.8em} 50\% layers, $\lambda=0.5$, $\mathfrak{B}=256$ & 83.51 & 80.03 & 75.24 & 79.45 & 0.099 & 83.21 & 80.04 & 75.02 & 79.28 & 0.098 & 83.07 & 79.32 & 73.29 & 78.35 & 0.118 \\
\addlinespace[1.5pt]
\hspace{0.8em} 50\% layers, $\lambda=1.0$, $\mathfrak{B}=128$ & 80.94 & 80.46 & 76.46 & 79.23 & 0.055 & 80.40 & 80.33 & 76.47 & 79.02 & 0.049 & 79.14 & 79.48 & 75.16 & 77.88 & 0.054 \\
\hspace{0.8em} 50\% layers, $\lambda=1.0$, $\mathfrak{B}=256$ & 80.52 & 78.68 & 76.71 & 78.61 & 0.047 & 79.85 & 78.48 & 76.84 & 78.37 & 0.038 & 78.60 & 77.67 & 75.46 & 77.22 & 0.040 \\
\bottomrule
\end{tabular}
}
\caption{nDCG@10 on FineWeb-PosQ across three embedding models and three
calibration variants. Columns within each variant group are the positional
groups (passage thirds), the harmonic mean across groups (HM), and the Position
Sensitivity Index (PSI). Baseline rows do not depend on the calibration variant
and are reported once.}
\label{tab:posq-comparison}
\end{table*}

\subsection{Efficient Implementation}
\label{sec:efficient-impl}
 
Calibration operates on post-softmax attention weights, requiring the encoder to run with
eager attention and materialize the full attention matrix
$\mathbf{A} \in \mathbb{R}^{B \times H \times Q \times K}$.
The original implementation
\citep{schuhmacher2026informationrepresentationfairnesslongdocument}
redistributes over this entire tensor.
 
Our implementation operates only on the attention rows that are modified, replaces 
per-position basket bookkeeping with a contiguous reduction, updates in place,
and reuses per-batch structure across calibrated layers. At 2{,}048 tokens and
batch size 8, it adds under 1\,MiB to peak memory and 6--9\% to encoding time over an
uncalibrated pass, compared with 5--7\,GiB and 126--219\% for the original
implementation. Appendix~\ref{app:implementation} gives details and the
correctness check.

\section{Experiments}

\subsection{Examined Calibration Configurations}
We evaluate calibration across a grid of four axes to characterize how each affects positional sensitivity and retrieval effectiveness, rather than to optimize for a specific benchmark.
Our choices for layer set and basket size follow the configurations examined by
\citet{schuhmacher2026informationrepresentationfairnesslongdocument} for representational fairness. In their experiments, calibrating the last layer 
already reduced positional bias, while calibrating the last 50\% of layers produced
a flatter representation profile across positions. We adopt these two layer
sets, $\mathfrak{L}^C \in \{\text{last layer},\ \text{last 50\%}\}$, and examine basket
sizes $\mathfrak{B} \in \{128, 256\}$. For calibration strength we examine
$\lambda = 1.0$ (full calibration, as in the original work) and $\lambda = 0.5$
(halfway between original and calibrated). We cross these with the three
calibration variants of Section~\ref{sec:selfattn}, giving 24 configurations per
model. The same grid is evaluated for all models, with no model-specific tuning, allowing us to identify configurations that perform consistently across architectures.

Figure~\ref{fig:schematic} illustrates the relationship between the basket
partition and the evaluation segments of FineWeb-PosQ. Basket size is chosen
independently of the downstream evaluation segments, which group queries by the position of their relevant information; different values of
$\mathfrak{B}$ therefore redistribute attention at different granularities
across the passages.

\subsection{Base Models}
We evaluate three dense retrieval models covering both common pooling architectures: gte-multilingual-base (M-GTE) \citep{zhang-etal-2024-mgte} and bge-m3-dense (BGE-M3) \citep{chen-etal-2024-m3} use \texttt{<s>}-pooling, and Qwen3-Embedding-0.6B (QWEN3-EMB) \citep{zhang2025qwen3embeddingadvancingtext} uses last-token pooling.
This distinction is important because the pooling token determines which attention query is calibrated: the first-token query for \texttt{<s>}-pooled models and the final-token query for last-token-pooled models.

Because positional attention patterns differ between models, we treat transfer across pooling architectures as an empirical question rather than assuming that calibration behaves similarly in both.

\subsection{Evaluation Metrics}
We report retrieval performance using nDCG@10 (Normalized Discounted Cumulative Gain at rank 10) \citep{10.1145/582415.582418} with binary relevance labels, following \citet{zeng-etal-2025-empirical}.

To summarize performance across positional groups, we report the harmonic mean (\textit{HM}):
\[
\text{HM} = \frac{n}{\sum_{i=1}^{n} \frac{1}{\text{nDCG@10}_i}},
\]
where $n$ is the number of positional groups and $\text{nDCG@10}_i$ is the score for group $i$. 
The harmonic mean penalizes low-performing groups more strongly than the arithmetic mean, providing a summary of retrieval effectiveness that gives greater weight to poorly performing positions.
We therefore use \textit{HM} as the primary summary metric for FineWeb-PosQ.

We additionally report the Position Sensitivity Index (PSI) introduced by \citet{zeng-etal-2025-empirical}, which captures the relative gap between the best- and worst-performing positional groups. 
Lower PSI indicates less positional variation, but is desirable only when retrieval effectiveness is preserved. We therefore interpret PSI together with \textit{HM} or aggregate nDCG@10 rather than as a standalone success criterion.

\subsection{Segment Embed Average Baseline}
As an additional baseline and intuitive reference point, we introduce \emph{Segment Embed Average \textit{(SEA)}}, which constructs passage embeddings by segmenting each passage with the semantic chunking library Chonkie \citep{chonkie2025}, embedding each chunk separately, and averaging the resulting representations. 
Motivated by findings on retrieval granularity \citep{chen-etal-2024-dense}, this baseline tests whether a simple segmentation-and-aggregation strategy can reduce positional dependence without modifying model internals.
Because averaging segment embeddings may lose information from the full passage, we use SEA mainly to contextualize the trade-off between positional sensitivity and retrieval effectiveness.

\begin{figure*}[t]
\centering
\includegraphics[width=0.98\textwidth]{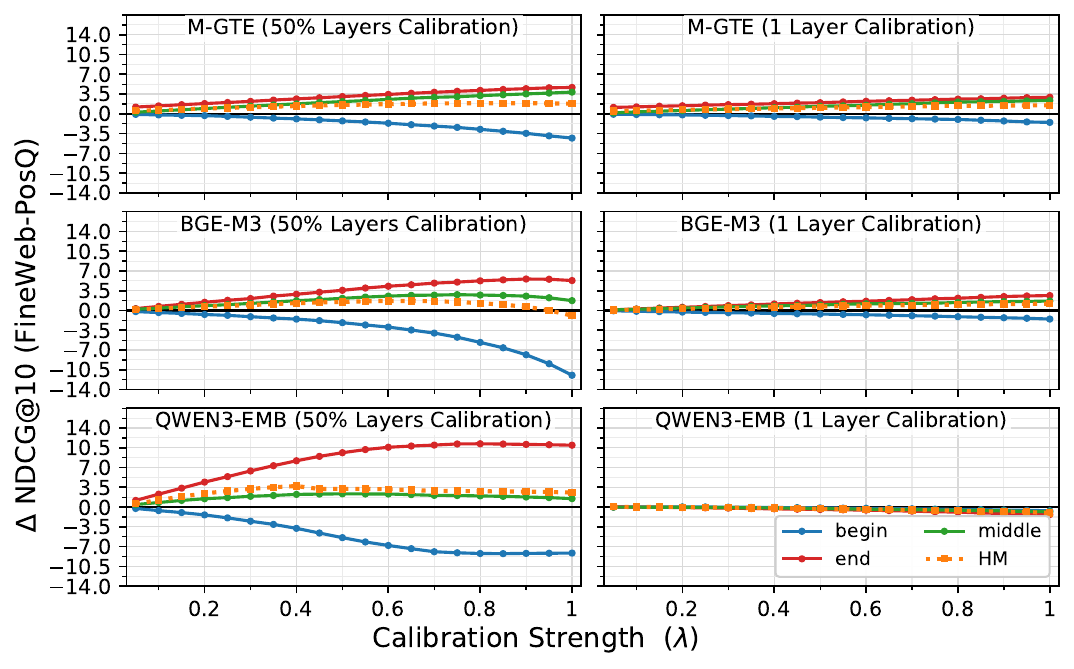} 
\caption{Absolute change in nDCG@10 on FineWeb-PosQ under \soft, relative to the uncalibrated baseline, as a function of calibration strength ($\lambda$). Other calibration variants behave analogously.}
\Description{A grid of six line plots, one row per model and two columns for
50\% layer calibration and last-layer calibration. Each plot shows the change
in nDCG@10 against $\lambda$ from 0 to 1 for the begin, middle and end groups
and the harmonic mean. In the left column the end curve rises and the begin
curve falls as $\lambda$ increases, with the harmonic mean rising early and
then flattening; the spread is widest for Qwen3-Embedding-0.6B. In the right
column all four curves stay close to zero across the whole range.}
\label{fig:ablation}
\end{figure*}

\section{Examination on FineWeb-PosQ}

Table~\ref{tab:posq-comparison} presents nDCG@10 across all calibration configurations on FineWeb-PosQ.

Calibration improves the harmonic mean under multiple configurations for all three models.
Calibration usually reduces nDCG@10 for the beginning group while improving the middle and end groups, reflecting the intended redistribution of attention away from early positions.
Across multiple configurations, these changes yield net harmonic-mean gains.
The best configuration improves the harmonic mean from 78.66 to 80.49 for gte-multilingual-base and  from 79.09 to 80.63 for bge-m3-dense, both under \soft; for Qwen3-Embedding-0.6B, it improves from 76.49 to 79.94 under \uniform.
For Qwen3-Embedding-0.6B, that setting also reduces the gap between the beginning and end groups: beginning performance decreases from 88.54 to 83.46, while end performance increases from 65.61 to 75.63.

The SEA baseline reduces PSI far more strongly but at a clear cost to overall effectiveness for the two \texttt{<s>}-pooled models (78.66 $\rightarrow$ 76.75 and 79.09 $\rightarrow$ 74.65); only for Qwen3-Embedding-0.6B, whose uncalibrated positional spread is largest, does it improve the harmonic mean.

\paragraph{Comparing redistribution variants}
The three variants differ in how the pooling token's own attention is treated, and the difference is largest where the intervention is strongest. At 50\% layer depth and $\lambda=1.0$, \uniform loses 1.03 harmonic-mean points for gte-multilingual-base and 1.88 for bge-m3-dense relative to the uncalibrated model, and \hard, which raises the pooling token's self-mass well above what this model allocates, loses 10.36 points for bge-m3-dense. \soft, which retains that mass, instead \emph{gains} 1.80 points for gte-multilingual-base at the same setting and limits the loss for bge-m3-dense to 0.91. Preserving the pooling token's self-mass therefore makes strong redistribution more robust: the same parameters that cost the other two variants up to 10.4 points remain usable under \soft.

\hard sharpens this. Because it fixes the pooling token's share at a constant, it
moves the two models in opposite directions: relative to \uniform, $w=0.3$ brings
gte-multilingual-base closer to its own allocation and pushes bge-m3-dense well
above it. The harmonic-mean changes follow the magnitude of this deviation rather than its direction ---
\hard gains 0.24 points over \uniform for gte-multilingual-base and loses 8.48
for bge-m3-dense. These results suggest that the problem is not whether
self-mass is increased or decreased, but whether its calibrated value departs substantially from the model's original allocation. \soft avoids such departures by construction. The effect is most pronounced at full
strength: at $\lambda=0.5$ all three variants stay within a point of each other
on both models.

The advantage of \soft is systematic rather than limited to the strongest setting. For
the two \texttt{<s>}-pooled models, \soft attains the highest harmonic mean in
all 16 model-configuration combinations and the
lowest PSI in 13 of 16; it also reduces sensitivity to $\lambda$, making
full-strength calibration substantially less costly. Qwen3-Embedding-0.6B, which
pools on the final token, shows no consistent advantage for \soft:
\uniform is marginally ahead on harmonic mean at moderate strength (79.94 against
79.67 at 50\% layers, $\lambda=0.5$), while \soft attains the lower PSI at
$\lambda=1.0$. Neither variant dominates consistently for this model.

\paragraph{Interaction between calibration strength and layer depth}
Calibration strength $\lambda$ and layer depth $\mathfrak{L}^C$ jointly determine the extent of modification to the pooling token's attention distribution. At $\lambda=0.5$, early-position performance is partly preserved while later positions improve moderately.
At $\lambda=1.0$, later positions often improve further, but early positions degrade more strongly.
Calibrating more layers amplifies this effect: last-layer calibration at $\lambda=1.0$ moves in the same direction as 50\% layer calibration at $\lambda=0.5$, but to a lesser extent.

The strongest intervention, 50\% layer calibration at $\lambda=1.0$, illustrates the cost of over-calibration. This configuration produces the flattest representation profile in \citet{schuhmacher2026informationrepresentationfairnesslongdocument}, but degrades retrieval under \uniform for both \texttt{<s>}-pooled models: the harmonic mean drops by 1.03 points for gte-multilingual-base (78.66 $\rightarrow$ 77.63) and 1.88 points for bge-m3-dense (79.09 $\rightarrow$ 77.21). Qwen3-Embedding-0.6B gains 2.74 points (76.49 $\rightarrow$ 79.23), but beginning-group performance drops by 7.60 points (88.54 $\rightarrow$ 80.94). As noted above, this degradation is not observed to the same extent under \soft, suggesting that overwriting the pooling token's self-mass contributes substantially to the cost of strong redistribution. 

\begin{table*}[t]
\centering
\newcommand{\imp}[1]{{\scriptsize\color{teal}\,#1}}
\newcommand{\degrade}[1]{{\scriptsize\color{red!70!black}\,#1}}
\begin{adjustbox}{max width=\textwidth}
\renewcommand{\arraystretch}{0.965}
\begin{tabular}{ll
  r@{\hspace{1pt}}l
  r@{\hspace{1pt}}l
  r@{\hspace{1pt}}l
  r@{\hspace{1pt}}l
  r@{\hspace{1pt}}l}
\toprule
\multirow{2}{*}{\textbf{Model}} & \multirow{2}{*}{\textbf{Indexing Method}} & \multicolumn{2}{c}{\textbf{Q1(512)}} & \multicolumn{2}{c}{\textbf{Q2(1024)}} & \multicolumn{2}{c}{\textbf{Q3(1536)}} & \multicolumn{2}{c}{\textbf{Q4(2048)}} & \multicolumn{2}{c}{\textbf{PosIR}} \\
&  & \multicolumn{2}{c}{\textbf{PSI}~$\downarrow$} & \multicolumn{2}{c}{\textbf{PSI}~$\downarrow$} & \multicolumn{2}{c}{\textbf{PSI}~$\downarrow$} & \multicolumn{2}{c}{\textbf{PSI}~$\downarrow$} & \multicolumn{2}{c}{\textbf{nDCG@10}~$\uparrow$} \\
\midrule
\multicolumn{12}{l}{\textit{Multilingual Retrieval (x$\rightarrow$x)}} \\
\midrule
\multirow{5}{*}{gte-multilingual-base}
  & No Calibration       & 0.21 & & 0.44 & & 0.56 & & 0.62 & & 47.37 & \\
  & Segment Embed Average        & 0.26 & \degrade{+0.05} & 0.18 & \imp{$-$0.26} & 0.25 & \imp{$-$0.31} & 0.30 & \imp{$-$0.32} & 42.86 & \degrade{$-$4.51} \\
  & \uniform & 0.16 & \imp{$-$0.05} & 0.37 & \imp{$-$0.07} & 0.47 & \imp{$-$0.09} & 0.52 & \imp{$-$0.10} & 48.51 & \imp{+1.14} \\
  & \soft & 0.15 & \imp{$-$0.06} & 0.35 & \imp{$-$0.09} & 0.45 & \imp{$-$0.11} & 0.50 & \imp{$-$0.12} & 49.63 & \imp{+2.26} \\
  & \hard & 0.16 & \imp{$-$0.05} & 0.37 & \imp{$-$0.07} & 0.47 & \imp{$-$0.09} & 0.52 & \imp{$-$0.10} & 49.41 & \imp{+2.04} \\
\addlinespace
\multirow{5}{*}{bge-m3-dense}
  & No Calibration       & 0.26 & & 0.33 & & 0.39 & & 0.36 & & 55.96 & \\
  & Segment Embed Average        & 0.25 & \imp{$-$0.01} & 0.21 & \imp{$-$0.12} & 0.27 & \imp{$-$0.12} & 0.35 & \imp{$-$0.01} & 45.52 & \degrade{$-$10.44} \\
  & \uniform & 0.15 & \imp{$-$0.11} & 0.23 & \imp{$-$0.10} & 0.27 & \imp{$-$0.12} & 0.24 & \imp{$-$0.12} & 56.42 & \imp{+0.46} \\
  & \soft & 0.16 & \imp{$-$0.10} & 0.22 & \imp{$-$0.11} & 0.27 & \imp{$-$0.12} & 0.22 & \imp{$-$0.14} & 56.56 & \imp{+0.60} \\
  & \hard & 0.17 & \imp{$-$0.09} & 0.23 & \imp{$-$0.10} & 0.28 & \imp{$-$0.11} & 0.24 & \imp{$-$0.12} & 56.29 & \imp{+0.33} \\
\midrule
\multicolumn{12}{l}{\textit{Cross-lingual Retrieval (x$\rightarrow$en)}} \\
\midrule
\multirow{5}{*}{gte-multilingual-base}
  & No Calibration       & 0.15 & & 0.32 & & 0.36 & & 0.43 & & 46.49 & \\
  & Segment Embed Average & 0.29 & \degrade{+0.14} & 0.18 & \imp{$-$0.14} & 0.31 & \imp{$-$0.05} & 0.27 & \imp{$-$0.16} & 38.20 & \degrade{$-$8.29} \\
  & \uniform & 0.14 & \imp{$-$0.01} & 0.23 & \imp{$-$0.09} & 0.26 & \imp{$-$0.10} & 0.31 & \imp{$-$0.12} & 46.46 & \degrade{$-$0.03} \\
  & \soft & 0.14 & \imp{$-$0.01} & 0.22 & \imp{$-$0.10} & 0.24 & \imp{$-$0.12} & 0.29 & \imp{$-$0.14} & 47.69 & \imp{+1.20} \\
  & \hard & 0.14 & \imp{$-$0.01} & 0.24 & \imp{$-$0.08} & 0.27 & \imp{$-$0.09} & 0.33 & \imp{$-$0.10} & 47.32 & \imp{+0.83} \\
\addlinespace
\multirow{5}{*}{bge-m3-dense}
  & No Calibration       & 0.28 & & 0.37 & & 0.41 & & 0.37 & & 51.23 & \\
  & Segment Embed Average        & 0.27 & \imp{$-$0.01} & 0.22 & \imp{$-$0.15} & 0.34 & \imp{$-$0.07} & 0.36 & \imp{$-$0.01} & 41.10 & \degrade{$-$10.13} \\
  & \uniform & 0.16 & \imp{$-$0.12} & 0.26 & \imp{$-$0.11} & 0.28 & \imp{$-$0.13} & 0.23 & \imp{$-$0.14} & 51.37 & \imp{+0.14} \\
  & \soft & 0.16 & \imp{$-$0.12} & 0.25 & \imp{$-$0.12} & 0.27 & \imp{$-$0.14} & 0.21 & \imp{$-$0.16} & 51.15 & \degrade{$-$0.08} \\
  & \hard & 0.17 & \imp{$-$0.11} & 0.27 & \imp{$-$0.10} & 0.28 & \imp{$-$0.13} & 0.22 & \imp{$-$0.15} & 50.98 & \degrade{$-$0.25} \\
\bottomrule
\end{tabular}
\end{adjustbox}
\caption{
Position Sensitivity Index (PSI)~$\downarrow$ and PosIR nDCG@10~$\uparrow$ of
gte-multilingual-base and bge-m3-dense, comparing the uncalibrated model,
Segment Embed Average, and the three calibration variants at
$\mathfrak{B}=128$, $\lambda=0.5$ and 50\% layer depth.
In both multilingual and cross-lingual retrieval (translated queries retrieving
English documents) settings, results are weighted-averaged across 31 domains and
then macro-averaged across 10 languages.
Deltas are relative to the \textit{No Calibration} setting.
}
\label{tab:psi_res}
\end{table*}

\subsection{Ablation: Calibration Strength}
To examine how calibration strength affects retrieval effectiveness---and where redistribution begins to hurt rather than help---we conduct a fine-grained ablation over $\lambda \in [0.05, 1.0]$ in increments of 0.05, with basket size fixed at $\mathfrak{B}=128$. Figure~\ref{fig:ablation} shows the change in nDCG@10 on FineWeb-PosQ under \soft relative to the uncalibrated baseline for each positional group, comparing 50\% layer calibration with last-layer calibration; the other redistribution variants behave analogously.

Increasing $\lambda$ improves nDCG@10 for the end group and degrades the beginning group across all three models, while the middle group shows a moderate upward trend. The harmonic mean rises early in every case, then flattens: for gte-multilingual-base and Qwen3-Embedding-0.6B it is essentially unchanged beyond $\lambda\!\approx\!0.5$ and $\lambda\!\approx\!0.4$ respectively, while for bge-m3-dense it peaks near $\lambda\!\approx\!0.6$ and falls back to the uncalibrated level by $\lambda=1.0$, as the beginning-group cost (approaching $-$11 points) overtakes the end-group gain. No model gains meaningfully beyond moderate strength, while bge-m3-dense loses its earlier gains by $\lambda=1.0$. This makes $\lambda=0.5$ a reasonable operating point across models.

The depth of calibration controls how quickly this trade-off appears. With 50\% layer calibration the curves are steep: for Qwen3-Embedding-0.6B, end-group gains exceed +10 nDCG@10 points while beginning-group losses approach $-$9. Last-layer calibration produces much smaller changes throughout and, for Qwen3-Embe\linebreak{}dding-0.6B, leaves all three positional groups essentially unchanged across the full range of $\lambda$. For this 28-layer last-token-pooled model, calibrating only the final layer has little effect on it's representation.

\subsection{Parameter Configuration for Large-Scale Evaluation}
Based on the parameter analysis and the $\lambda$ sweep, we select a common operating point across models and variants: $\mathfrak{B}=128$, $\lambda=0.5$, with 50\% layer calibration. Smaller baskets give slightly higher harmonic means than $\mathfrak{B}=256$ throughout, and every model reaches its best or near-best harmonic mean at this strength under at least one variant. We fix these parameters for the large-scale evaluation that follows and vary only the redistribution variant, so that differences between variants reflect their treatment of the pooling token's self-mass.

\section{Position-Aware IR Benchmark}

To test whether the configuration selected on FineWeb-PosQ  generalizes beyond that benchmark, we evaluate it on PosIR \citep{zeng2026posir}, which spans 10 languages, 31 domains, and four document-length quartiles (Q1: 512 tokens to Q4: 2048 tokens). PosIR's length-controlled bucketing isolates position bias from length-induced degradation, allowing us to test whether calibration reduces position sensitivity across languages, domains, and document lengths not used for configuration selection.
Due to the computational cost of PosIR, we evaluate two of the three models from our parameter exploration, gte-multilingual-base and bge-m3-dense; the calibrated PosIR runs alone required over 1{,}000 A100 and 3{,}500 H100 GPU-hours.
We evaluate both models in the multilingual (x$\rightarrow$x) and cross-lingual (x$\rightarrow$en) settings, applying calibration at indexing time with $\mathfrak{B}=128$, $\lambda=0.5$, and 50\% layer depth. All three redistribution variants are evaluated at these parameters, so differences between them reflect their treatment of the pooling token's self-mass. Following PosIR's protocol, we compute a weighted average across the 31 domains within each language and then macro-average across the 10 languages. Per-language results for all
variants are given in Appendix Tables~\ref{tab:appendix_multilingual_gte}--\ref{tab:appendix_crosslingual_bgem3}.

FineWeb-PosQ and PosIR summarize positional effects differently. In FineWeb-PosQ, positional groups are defined by the location of the answer or reference span within a passage, and we report nDCG@10 separately for these groups. In PosIR, following the benchmark protocol, aggregate nDCG@10 is reported across all queries, while PSI is computed within document-length quartiles from position-conditioned performance. We therefore use PosIR primarily as a generalization test for position sensitivity reduction, with aggregate nDCG@10 serving as an effectiveness check.

\paragraph{PSI is reduced consistently.} Table~\ref{tab:psi_res} shows that
calibration reduces PSI in all 16 length-quartile $\times$ model $\times$
retrieval-setting combinations for every variant.
For gte-multilingual-base, the reductions tend to increase with document length: in the multilingual
setting, PSI drops by 0.05 at Q1 (512 tokens) and by
0.10 at Q4 (2048 tokens). For bge-m3-dense, reductions are roughly stable at
0.10--0.12 across quartiles.
The cross-lingual setting shows similar reductions, reaching 0.16
for bge-m3-dense at Q4.
The largest absolute reductions occur in the longer quartiles, where
the uncalibrated baselines also tend to show greater position sensitivity.
The three variants behave similarly here, with \soft marginally
ahead: it attains the lowest or tied-lowest PSI in 15 of the 16 combinations,
though its margin over \uniform rarely exceeds 0.02.

\paragraph{Aggregate retrieval effectiveness is preserved or improved.} Aggregate nDCG@10 is largely preserved or improved alongside these reductions. Here the variants separate more clearly than they do on PSI. \soft improves aggregate nDCG@10 in three of the four model $\times$ setting combinations, with the largest gain among the variants in each: +2.26 and +0.60 points in the multilingual setting for gte-multilingual-base and bge-m3-dense, and +1.20 for gte-multilingual-base in the cross-lingual setting, against +1.14, +0.46 and $-$0.03 for \uniform. The exception is bge-m3-dense in the cross-lingual setting, where \soft costs 0.08 points while \uniform gains 0.14 --- a difference of 0.22 points, in the row where \soft holds the lowest PSI at every quartile. Overall, retaining the pooling token's self-mass provides the best balance between positional stability and retrieval effectiveness.

\paragraph{The trade-off is not automatic.} Segment Embed Average shows that reducing PSI does not necessarily preserve retrieval effectiveness: the baseline reaches similar PSI reductions only by sacrificing effectiveness, lowering aggregate nDCG@10 by 4.51--10.44 points (e.g.\ $-$10.44 for bge-m3-dense in the multilingual setting). Calibration achieves these reductions while largely preserving or improving nDCG@10.

\section{NanoBEIR: General Retrieval Effectiveness}

Calibration targets position bias, but should preserve general retrieval effectiveness. We therefore evaluate all
three variants on NanoBEIR \citep{nanobeir2024}, a set of reduced versions of the
13 BEIR datasets \citep{thakur2021beir} sampled to 50 queries each. Using the same parameters as above, we report averages over the 13
datasets in Table~\ref{tab:nanobeir_main}, with
per-dataset nDCG@10 and Recall@10 in Appendix Tables~\ref{tab:nanobeir_ndcg}
and~\ref{tab:nanobeir_recall}. NanoBEIR documents are short relative
to PosIR, so most fall within one or two baskets and calibration has little attention mass to
redistribute; we therefore expect only small changes.

Aggregate nDCG@10 changes by at most 0.15 points for gte-multilingual-base. For bge-m3-dense, \soft differs from the uncalibrated model by 0.27 points, compared with 0.72 under \uniform, making \soft the closest variant to the uncalibrated baseline. Recall@10 is similarly stable, with \soft achieving the highest value for bge-m3-dense.

We additionally encode each document after randomly reshuffling its sentences, disrupting the document's original positional structure. All methods lose effectiveness under reshuffling, but the loss is smallest for \soft in both models (0.75 and 1.78 nDCG@10 points, vs. 0.99 and 1.88 uncalibrated), suggesting that \soft is less sensitive to changes in content position.

\begin{table}[t]
\centering
\setlength{\tabcolsep}{5pt}
\renewcommand{\arraystretch}{0.95}
\footnotesize
\begin{tabular}{lcccc}
\toprule
 & \multicolumn{2}{c}{\textbf{nDCG@10}} & \multicolumn{2}{c}{\textbf{Recall@10}} \\
\cmidrule(lr){2-3}\cmidrule(lr){4-5}
 & Original & Shuffled & Original & Shuffled \\
\midrule
\multicolumn{5}{l}{\textbf{gte-multilingual-base}} \\
No Calibration & 62.36 & 61.37 & 65.47 & 65.08 \\
\uniform & 62.27 & 61.18 & 65.75 & 65.18 \\
\soft & 62.21 & 61.46 & 65.30 & 65.12 \\
\hard & 62.35 & 61.26 & 65.86 & 65.32 \\
\midrule
\multicolumn{5}{l}{\textbf{bge-m3-dense}} \\
No Calibration & 60.59 & 58.71 & 64.32 & 63.31 \\
\uniform & 59.87 & 57.81 & 63.75 & 62.38 \\
\soft & 60.32 & 58.54 & 64.72 & 63.24 \\
\hard & 60.31 & 58.29 & 64.41 & 62.69 \\
\bottomrule
\end{tabular}
\caption{NanoBEIR averages over the 13 datasets; \textit{Shuffled} encodes each document with its sentences
randomly reshuffled before embedding. Values are $\times 100$.}
\label{tab:nanobeir_main}
\end{table}


\section{Conclusions}
We studied inference-time attention calibration as a mitigation for positional
bias in dense retrieval and extended it in two directions. A strength coefficient
$\lambda \in [0,1]$ interpolates between the original and calibrated attention,
making the intervention controllable rather than fixed. \soft preserves the pooling
token's self-mass instead of overwriting it with $1/n_{\mathrm{baskets}}$,
so the redistribution budget follows the model at each layer. Our efficient implementation reduces peak calibration memory overhead from 5–7\,GiB to under 1\,MiB, making large-scale evaluation practical.

Moderate calibration provides a better retrieval trade-off
than the full equalization: stronger settings improve
late-position performance at a greater cost to early positions. This trade-off also depends on how attention is redistributed. Our results suggest that large deviations from the pooling
token's original self-mass contribute substantially to the cost of strong calibration: harmonic mean decreases by up to 1.9 points under \uniform and by up to 10.4 points when self-mass is fixed independently
of the model. \soft avoids such deviations by construction  and provides the best overall balance between position sensitivity and retrieval effectiveness on PosIR.

The selected configuration transfers beyond FineWeb-PosQ: on PosIR, it
reduces position sensitivity in every evaluated combination of length quartile,
model, and retrieval setting across 10 languages and 31 domains,
while largely preserving or improving  aggregate retrieval effectiveness.

We therefore adopt \soft as the default in our implementation: it performs best for
the evaluated \texttt{<s>}-pooled models and remains competitive for the last-token-pooled model. With $\mathfrak{B}=128$, $\lambda=0.5$ and 50\% layer
depth, the same configuration applies across all four document-length quartiles and ten
languages of PosIR, in both multilingual and cross-lingual retrieval, without
per-model or per-language tuning. Calibration improves retrieval of information at later passage positions without retraining or query-time latency, while leaving general retrieval effectiveness essentially unchanged.

\section*{Ethics and Privacy Statement}

This work uses publicly released models and benchmarks (FineWeb-PosQ, PosIR,
NanoBEIR). We collected no new data and ran no study involving human
participants. Calibration operates on a model's own attention weights at
indexing time and does not read or store anything about users or their queries.

The fairness we address is positional: whether content is reflected in a
document's embedding regardless of where in the document it appears. It is not
fairness with respect to people or social groups, and calibration does not
mitigate demographic or linguistic bias. The main risk we see is
overcorrection, since strong calibration improves late-position retrieval at a
cost to early positions: tuning on a positional-spread metric alone can degrade
retrieval while appearing to make it fairer. We therefore report PSI together
with retrieval effectiveness throughout.

\begin{acks}
This research is conducted under the project \textit{Impresso -- Media Monitoring of the Past II. Beyond Borders: Connecting Historical Newspapers and Radio}. Impresso is a research project funded by the Swiss National Science Foundation (SNSF 213585) and the Luxembourg National Research Fund (17498891).
\end{acks}

\bibliographystyle{ACM-Reference-Format}
\bibliography{sample-base}
\appendix

\section{Implementation Details}
\label{app:implementation}
\label{sec:appendix}

The attention calibration is implemented via \texttt{nnsight} \citep{fiotto-kaufman2025nnsight}, an interpretability library that provides hooks into PyTorch modules during the forward pass. By setting \texttt{attn\_implementation="eager"}, we intercept the post-softmax attention weight tensors $\mathbf{A} \in \mathbb{R}^{B \times H \times Q \times K}$ at specified layers and apply basket-level equalization in-place before the forward pass continues. This operates entirely at inference time with no gradient computation.

The calibration modifies only the query row corresponding to the pooling token, leaving all other query rows unchanged. For \texttt{<s>}-pooled models (also called \texttt{[CLS]}-pooled models), this is the first token's query row; for last-token-pooled models, this is the final token's query row \texttt{</s>} (also called \texttt{[EOS]}). The \texttt{<s>} and \texttt{</s>} tokens can optionally be isolated into their own baskets to prevent special-token attention from being redistributed across content baskets. Table~\ref{tab:calib-config} summarizes the default calibration configuration for each validated model, alongside the generic defaults for each pooling architecture.

\paragraph{Efficient implementation.}
The original implementation expresses basket membership as a position-to-basket index, accumulates basket sums with
\texttt{scatter\_add}, redistributes them with \texttt{gather}, and allocates a
further $B \times H \times Q \times K$ tensor at each subsequent step. We
exploit instead the fact that baskets are contiguous blocks of fixed size
$\mathfrak{B}$: after aligning each passage's content span to a common offset,
reshaping the key axis to $(n_{\mathrm{baskets}}, \mathfrak{B})$ makes the
basket sums a reduction along the trailing axis and the renormalization a
broadcast division, removing the index tensor and both indexed operations.
Alignment is per batch item, because the content span begins at a different
offset under left padding. For \texttt{<s>}- and \texttt{</s>}-pooled models
only the pooled query row is extracted, calibrated and written back. Batch geometry that depends only on the attention
mask is computed once per batch rather than once per calibrated layer. At matched batch size and
2{,}048 tokens, ours adds 6\% (gte-multilingual-base) and 9\% (bge-m3-dense) to
encoding time, against 126\% and 219\% for the original implementation; the
spread between
models tracks the share of the forward pass that is calibrated. Throughput
gained from the larger batches the smaller footprint permits is not counted.
 
\paragraph{Correctness.}
We verify our implementation against the original implementation on 25{,}920 randomized cases, spanning both pooling
schemes, both padding sides, batch sizes 1--8, basket counts 1--46, all three
redistribution variants, and \texttt{float32} and \texttt{float16}; the two
agree to $2.4 \times 10^{-7}$ in \texttt{float32}. The residual difference is
reduction order: a contiguous reduction associates the summands differently from
\texttt{scatter\_add}, which on CUDA accumulates through atomics and is itself
non-deterministic across runs, so bit-identical agreement is attainable in
neither direction. On the two evaluated models in \texttt{float16}, embeddings
from the two implementations differ by at most $7.3 \times 10^{-4}$ per
component, at a cosine similarity of $1.000$.

\begin{table}[h]
\centering
\footnotesize
\setlength{\tabcolsep}{4pt}
\begin{tabular}{lccc}
\toprule
\multirow{2}{*}{\textbf{Model / Architecture}} & \multirow{2}{*}{\shortstack{\textbf{Pooling /}\\\textbf{calib. token}}} & \multicolumn{2}{c}{\textbf{Isolated}} \\
\cmidrule(lr){3-4}
 & & \texttt{<s>} & \texttt{</s>} \\
\midrule
\multicolumn{4}{l}{\textit{Defaults}} \\
\quad \texttt{<s>}-pooled & \texttt{<s>} & \cmark & --- \\
\quad \texttt{</s>}-pooled & \texttt{</s>} & \cmark & \cmark \\
\midrule
\multicolumn{4}{l}{\textit{Examined models}} \\
\quad gte-multilingual-base & \texttt{<s>} & \cmark & --- \\
\quad bge-m3-dense & \texttt{<s>} & \cmark & --- \\
\quad Qwen3-Embedding-0.6B & \texttt{</s>} & \cmark & \cmark \\
\bottomrule
\end{tabular}
\caption{Calibration configuration per model. The pooling token is also the
calibrated query row. Isolated columns indicate which special tokens are placed
in their own basket.}
\label{tab:calib-config}
\end{table}

\section{Extended Tables}

\begin{table*}[t]
\centering
\setlength{\tabcolsep}{4pt}
\renewcommand{\arraystretch}{0.85}
\providecommand{\imp}[1]{{\scriptsize\color{teal}\,#1}}
\providecommand{\degrade}[1]{{\scriptsize\color{red!70!black}\,#1}}
\small
\begin{adjustbox}{max width=\textwidth}
\begin{tabular}{l
  r@{\hspace{1pt}}l
  r@{\hspace{1pt}}l
  r@{\hspace{1pt}}l
  r@{\hspace{1pt}}l
  r@{\hspace{1pt}}l
  r@{\hspace{1pt}}l}
\toprule
\textbf{Multilingual} & \multicolumn{2}{c}{} & \multicolumn{10}{c}{\textbf{nDCG@10}~$\uparrow$} \\
\cmidrule(lr){4-13}
\textbf{Retrieval} & \multicolumn{2}{c}{\textbf{PSI}~$\downarrow$} & \multicolumn{2}{c}{\textbf{Q1}} & \multicolumn{2}{c}{\textbf{Q2}} & \multicolumn{2}{c}{\textbf{Q3}} & \multicolumn{2}{c}{\textbf{Q4}} & \multicolumn{2}{c}{\textbf{Overall}~$\uparrow$} \\
\midrule[\heavyrulewidth]
\multicolumn{13}{l}{\footnotesize\textbf{gte-multilingual-base}\quad\textit{No Calibration}} \\
Arabic & 0.52 &   & 49.55 &   & 35.75 &   & 24.71 &   & 18.41 &   & 34.08 &   \\
Chinese & 0.26 &   & 61.52 &   & 52.22 &   & 45.26 &   & 39.21 &   & 51.52 &   \\
German & 0.44 &   & 61.47 &   & 49.87 &   & 40.32 &   & 32.10 &   & 47.96 &   \\
English & 0.19 &   & 73.24 &   & 62.56 &   & 54.46 &   & 48.46 &   & 61.59 &   \\
French & 0.41 &   & 64.12 &   & 50.17 &   & 41.32 &   & 35.11 &   & 49.59 &   \\
Italian & 0.45 &   & 61.56 &   & 49.25 &   & 40.12 &   & 31.76 &   & 47.63 &   \\
Korean & 0.48 &   & 54.06 &   & 39.15 &   & 27.32 &   & 20.19 &   & 37.54 &   \\
Portuguese & 0.42 &   & 62.82 &   & 50.84 &   & 41.18 &   & 32.57 &   & 48.94 &   \\
Russian & 0.46 &   & 59.86 &   & 46.55 &   & 36.08 &   & 28.78 &   & 44.89 &   \\
Spanish & 0.42 &   & 64.62 &   & 51.54 &   & 41.30 &   & 33.50 &   & 49.95 &   \\
\textbf{Avg.} & 0.40 &   & 61.28 &   & 48.79 &   & 39.21 &   & 32.01 &   & 47.37 &   \\
\cmidrule{1-13}
\multicolumn{13}{l}{\footnotesize\textit{\uniform}} \\
Arabic & 0.44 & \imp{$-$0.08} & 50.31 & \imp{+0.76} & 37.47 & \imp{+1.72} & 25.44 & \imp{+0.73} & 17.86 & \degrade{$-$0.55} & 35.20 & \imp{+1.12} \\
Chinese & 0.14 & \imp{$-$0.12} & 63.23 & \imp{+1.71} & 53.93 & \imp{+1.71} & 45.84 & \imp{+0.58} & 38.61 & \degrade{$-$0.60} & 52.36 & \imp{+0.84} \\
German & 0.34 & \imp{$-$0.10} & 61.93 & \imp{+0.46} & 51.51 & \imp{+1.64} & 41.46 & \imp{+1.14} & 32.88 & \imp{+0.78} & 49.24 & \imp{+1.28} \\
English & 0.10 & \imp{$-$0.09} & 74.27 & \imp{+1.03} & 63.85 & \imp{+1.29} & 55.13 & \imp{+0.67} & 48.24 & \degrade{$-$0.22} & 62.61 & \imp{+1.02} \\
French & 0.32 & \imp{$-$0.09} & 64.81 & \imp{+0.69} & 52.50 & \imp{+2.33} & 43.05 & \imp{+1.73} & 36.39 & \imp{+1.28} & 51.37 & \imp{+1.78} \\
Italian & 0.36 & \imp{$-$0.09} & 62.16 & \imp{+0.60} & 51.20 & \imp{+1.95} & 41.49 & \imp{+1.37} & 32.82 & \imp{+1.06} & 49.12 & \imp{+1.49} \\
Korean & 0.37 & \imp{$-$0.11} & 54.24 & \imp{+0.18} & 39.71 & \imp{+0.56} & 27.03 & \degrade{$-$0.29} & 18.96 & \degrade{$-$1.23} & 37.79 & \imp{+0.25} \\
Portuguese & 0.34 & \imp{$-$0.08} & 63.54 & \imp{+0.72} & 52.72 & \imp{+1.88} & 42.29 & \imp{+1.11} & 33.32 & \imp{+0.75} & 50.33 & \imp{+1.39} \\
Russian & 0.37 & \imp{$-$0.09} & 60.65 & \imp{+0.79} & 48.02 & \imp{+1.47} & 36.68 & \imp{+0.60} & 28.50 & \degrade{$-$0.28} & 45.90 & \imp{+1.01} \\
Spanish & 0.33 & \imp{$-$0.09} & 65.10 & \imp{+0.48} & 53.31 & \imp{+1.77} & 42.60 & \imp{+1.30} & 33.92 & \imp{+0.42} & 51.21 & \imp{+1.26} \\
\textbf{Avg.} & 0.31 & \imp{$-$0.09} & 62.02 & \imp{+0.74} & 50.42 & \imp{+1.63} & 40.10 & \imp{+0.89} & 32.15 & \imp{+0.14} & 48.51 & \imp{+1.14} \\
\cmidrule{1-13}
\multicolumn{13}{l}{\footnotesize\textit{\soft}} \\
Arabic & 0.41 & \imp{$-$0.11} & 50.74 & \imp{+1.19} & 38.49 & \imp{+2.74} & 26.66 & \imp{+1.95} & 19.38 & \imp{+0.97} & 36.19 & \imp{+2.11} \\
Chinese & 0.12 & \imp{$-$0.14} & 63.08 & \imp{+1.56} & 55.14 & \imp{+2.92} & 48.52 & \imp{+3.26} & 42.09 & \imp{+2.88} & 53.84 & \imp{+2.32} \\
German & 0.32 & \imp{$-$0.12} & 61.87 & \imp{+0.40} & 52.36 & \imp{+2.49} & 42.72 & \imp{+2.40} & 34.87 & \imp{+2.77} & 50.13 & \imp{+2.17} \\
English & 0.08 & \imp{$-$0.11} & 74.15 & \imp{+0.91} & 65.08 & \imp{+2.52} & 57.68 & \imp{+3.22} & 52.03 & \imp{+3.57} & 64.14 & \imp{+2.55} \\
French & 0.30 & \imp{$-$0.11} & 64.71 & \imp{+0.59} & 53.48 & \imp{+3.31} & 44.86 & \imp{+3.54} & 38.40 & \imp{+3.29} & 52.38 & \imp{+2.79} \\
Italian & 0.35 & \imp{$-$0.10} & 62.15 & \imp{+0.59} & 52.11 & \imp{+2.86} & 43.08 & \imp{+2.96} & 34.79 & \imp{+3.03} & 50.08 & \imp{+2.45} \\
Korean & 0.35 & \imp{$-$0.13} & 54.35 & \imp{+0.29} & 40.93 & \imp{+1.78} & 28.37 & \imp{+1.05} & 20.46 & \imp{+0.27} & 38.74 & \imp{+1.20} \\
Portuguese & 0.32 & \imp{$-$0.10} & 63.47 & \imp{+0.65} & 53.62 & \imp{+2.78} & 43.83 & \imp{+2.65} & 35.36 & \imp{+2.79} & 51.27 & \imp{+2.33} \\
Russian & 0.35 & \imp{$-$0.11} & 60.52 & \imp{+0.66} & 49.18 & \imp{+2.63} & 39.07 & \imp{+2.99} & 31.15 & \imp{+2.37} & 47.18 & \imp{+2.29} \\
Spanish & 0.31 & \imp{$-$0.11} & 65.18 & \imp{+0.56} & 54.49 & \imp{+2.95} & 44.24 & \imp{+2.94} & 36.25 & \imp{+2.75} & 52.37 & \imp{+2.42} \\
\textbf{Avg.} & 0.29 & \imp{$-$0.11} & 62.02 & \imp{+0.74} & 51.49 & \imp{+2.70} & 41.90 & \imp{+2.69} & 34.48 & \imp{+2.47} & 49.63 & \imp{+2.26} \\
\cmidrule{1-13}
\multicolumn{13}{l}{\footnotesize\textit{\hard}} \\
Arabic & 0.46 & \imp{$-$0.06} & 49.49 & \degrade{$-$0.06} & 38.80 & \imp{+3.05} & 28.17 & \imp{+3.46} & 20.84 & \imp{+2.43} & 36.39 & \imp{+2.31} \\
Chinese & 0.14 & \imp{$-$0.12} & 62.26 & \imp{+0.74} & 54.15 & \imp{+1.93} & 47.74 & \imp{+2.48} & 41.94 & \imp{+2.73} & 53.13 & \imp{+1.61} \\
German & 0.36 & \imp{$-$0.08} & 60.96 & \degrade{$-$0.51} & 52.15 & \imp{+2.28} & 43.53 & \imp{+3.21} & 35.68 & \imp{+3.58} & 50.02 & \imp{+2.06} \\
English & 0.10 & \imp{$-$0.09} & 73.39 & \imp{+0.15} & 64.12 & \imp{+1.56} & 56.68 & \imp{+2.22} & 50.55 & \imp{+2.09} & 63.11 & \imp{+1.52} \\
French & 0.34 & \imp{$-$0.07} & 63.86 & \degrade{$-$0.26} & 53.15 & \imp{+2.98} & 45.14 & \imp{+3.82} & 38.72 & \imp{+3.61} & 52.08 & \imp{+2.49} \\
Italian & 0.37 & \imp{$-$0.08} & 61.19 & \degrade{$-$0.37} & 51.93 & \imp{+2.68} & 43.75 & \imp{+3.63} & 35.88 & \imp{+4.12} & 50.00 & \imp{+2.37} \\
Korean & 0.40 & \imp{$-$0.08} & 53.30 & \degrade{$-$0.76} & 41.01 & \imp{+1.86} & 30.03 & \imp{+2.71} & 22.40 & \imp{+2.21} & 39.05 & \imp{+1.51} \\
Portuguese & 0.36 & \imp{$-$0.06} & 62.58 & \degrade{$-$0.24} & 53.42 & \imp{+2.58} & 44.71 & \imp{+3.53} & 36.48 & \imp{+3.91} & 51.26 & \imp{+2.32} \\
Russian & 0.38 & \imp{$-$0.08} & 59.61 & \degrade{$-$0.25} & 48.87 & \imp{+2.32} & 39.30 & \imp{+3.22} & 31.64 & \imp{+2.86} & 46.89 & \imp{+2.00} \\
Spanish & 0.34 & \imp{$-$0.08} & 64.25 & \degrade{$-$0.37} & 54.00 & \imp{+2.46} & 44.76 & \imp{+3.46} & 37.08 & \imp{+3.58} & 52.13 & \imp{+2.18} \\
\textbf{Avg.} & 0.33 & \imp{$-$0.07} & 61.09 & \degrade{$-$0.19} & 51.16 & \imp{+2.37} & 42.38 & \imp{+3.17} & 35.12 & \imp{+3.11} & 49.41 & \imp{+2.04} \\
\cmidrule{1-13}
\multicolumn{13}{l}{\footnotesize\textit{Segment Embed Average}} \\
Arabic & 0.20 & \imp{$-$0.32} & 30.59 & \degrade{$-$18.96} & 32.76 & \degrade{$-$2.99} & 30.40 & \imp{+5.69} & 28.72 & \imp{+10.31} & 30.99 & \degrade{$-$3.09} \\
Chinese & 0.09 & \imp{$-$0.17} & 47.54 & \degrade{$-$13.98} & 45.08 & \degrade{$-$7.14} & 46.22 & \imp{+0.96} & 45.94 & \imp{+6.73} & 45.95 & \degrade{$-$5.57} \\
German & 0.21 & \imp{$-$0.23} & 43.67 & \degrade{$-$17.80} & 45.50 & \degrade{$-$4.37} & 42.68 & \imp{+2.36} & 40.50 & \imp{+8.40} & 43.61 & \degrade{$-$4.35} \\
English & 0.20 & \degrade{+0.01} & 58.26 & \degrade{$-$14.98} & 56.20 & \degrade{$-$6.36} & 51.67 & \degrade{$-$2.79} & 49.58 & \imp{+1.12} & 55.06 & \degrade{$-$6.53} \\
French & 0.21 & \imp{$-$0.20} & 46.69 & \degrade{$-$17.43} & 47.78 & \degrade{$-$2.39} & 44.86 & \imp{+3.54} & 41.76 & \imp{+6.65} & 45.97 & \degrade{$-$3.62} \\
Italian & 0.18 & \imp{$-$0.27} & 44.09 & \degrade{$-$17.47} & 45.28 & \degrade{$-$3.97} & 43.29 & \imp{+3.17} & 40.35 & \imp{+8.59} & 43.82 & \degrade{$-$3.81} \\
Korean & 0.26 & \imp{$-$0.22} & 33.82 & \degrade{$-$20.24} & 33.99 & \degrade{$-$5.16} & 31.43 & \imp{+4.11} & 29.60 & \imp{+9.41} & 32.79 & \degrade{$-$4.75} \\
Portuguese & 0.20 & \imp{$-$0.22} & 45.71 & \degrade{$-$17.11} & 46.51 & \degrade{$-$4.33} & 43.70 & \imp{+2.52} & 40.81 & \imp{+8.24} & 44.87 & \degrade{$-$4.07} \\
Russian & 0.25 & \imp{$-$0.21} & 40.73 & \degrade{$-$19.13} & 40.73 & \degrade{$-$5.82} & 38.17 & \imp{+2.09} & 35.80 & \imp{+7.02} & 39.55 & \degrade{$-$5.34} \\
Spanish & 0.21 & \imp{$-$0.21} & 47.94 & \degrade{$-$16.68} & 47.12 & \degrade{$-$4.42} & 44.41 & \imp{+3.11} & 41.46 & \imp{+7.96} & 46.04 & \degrade{$-$3.91} \\
\textbf{Avg.} & 0.20 & \imp{$-$0.20} & 43.90 & \degrade{$-$17.38} & 44.10 & \degrade{$-$4.69} & 41.68 & \imp{+2.48} & 39.45 & \imp{+7.44} & 42.86 & \degrade{$-$4.50} \\
\bottomrule
\end{tabular}
\end{adjustbox}
\caption{Multilingual retrieval with \textbf{gte-multilingual-base}. Q1--Q4 denote nDCG@10 by query quartile; Overall is nDCG@10 across all queries. nDCG@10 values are $\times 100$. Deltas are relative to \textit{No Calibration}; \textcolor{teal}{teal} indicates improvement, \textcolor{red!70!black}{red} indicates degradation.}
\label{tab:appendix_multilingual_gte}
\end{table*}

\begin{table*}[t]
\centering
\setlength{\tabcolsep}{4pt}
\renewcommand{\arraystretch}{0.85}
\providecommand{\imp}[1]{{\scriptsize\color{teal}\,#1}}
\providecommand{\degrade}[1]{{\scriptsize\color{red!70!black}\,#1}}
\small
\begin{adjustbox}{max width=\textwidth}
\begin{tabular}{l
  r@{\hspace{1pt}}l
  r@{\hspace{1pt}}l
  r@{\hspace{1pt}}l
  r@{\hspace{1pt}}l
  r@{\hspace{1pt}}l
  r@{\hspace{1pt}}l}
\toprule
\textbf{Multilingual} & \multicolumn{2}{c}{} & \multicolumn{10}{c}{\textbf{nDCG@10}~$\uparrow$} \\
\cmidrule(lr){4-13}
\textbf{Retrieval} & \multicolumn{2}{c}{\textbf{PSI}~$\downarrow$} & \multicolumn{2}{c}{\textbf{Q1}} & \multicolumn{2}{c}{\textbf{Q2}} & \multicolumn{2}{c}{\textbf{Q3}} & \multicolumn{2}{c}{\textbf{Q4}} & \multicolumn{2}{c}{\textbf{Overall}~$\uparrow$} \\
\midrule[\heavyrulewidth]
\multicolumn{13}{l}{\footnotesize\textbf{bge-m3-dense}\quad\textit{No Calibration}} \\
Arabic & 0.39 &   & 52.10 &   & 46.04 &   & 41.96 &   & 39.49 &   & 45.07 &   \\
Chinese & 0.25 &   & 65.14 &   & 55.66 &   & 51.13 &   & 47.83 &   & 56.54 &   \\
German & 0.25 &   & 67.05 &   & 58.84 &   & 53.54 &   & 49.84 &   & 58.36 &   \\
English & 0.22 &   & 73.37 &   & 64.38 &   & 59.47 &   & 54.04 &   & 63.96 &   \\
French & 0.29 &   & 67.51 &   & 57.51 &   & 53.11 &   & 49.37 &   & 57.75 &   \\
Italian & 0.29 &   & 67.28 &   & 57.68 &   & 52.53 &   & 48.33 &   & 57.45 &   \\
Korean & 0.31 &   & 59.74 &   & 49.48 &   & 44.70 &   & 41.48 &   & 49.62 &   \\
Portuguese & 0.30 &   & 66.91 &   & 58.83 &   & 53.82 &   & 49.50 &   & 58.07 &   \\
Russian & 0.29 &   & 63.07 &   & 54.60 &   & 50.05 &   & 46.39 &   & 54.29 &   \\
Spanish & 0.27 &   & 68.00 &   & 58.70 &   & 53.75 &   & 49.92 &   & 58.48 &   \\
\textbf{Avg.} & 0.29 &   & 65.02 &   & 56.17 &   & 51.40 &   & 47.62 &   & 55.96 &   \\
\cmidrule{1-13}
\multicolumn{13}{l}{\footnotesize\textit{\uniform}} \\
Arabic & 0.19 & \imp{$-$0.19} & 56.63 & \imp{+4.54} & 46.96 & \imp{+0.93} & 40.41 & \degrade{$-$1.55} & 36.92 & \degrade{$-$2.57} & 46.42 & \imp{+1.34} \\
Chinese & 0.12 & \imp{$-$0.14} & 68.53 & \imp{+3.39} & 56.17 & \imp{+0.51} & 49.94 & \degrade{$-$1.19} & 44.87 & \degrade{$-$2.95} & 56.65 & \imp{+0.11} \\
German & 0.10 & \imp{$-$0.15} & 69.90 & \imp{+2.85} & 58.39 & \degrade{$-$0.46} & 51.32 & \degrade{$-$2.22} & 46.28 & \degrade{$-$3.56} & 58.30 & \degrade{$-$0.07} \\
English & 0.07 & \imp{$-$0.16} & 76.36 & \imp{+2.99} & 64.87 & \imp{+0.49} & 57.48 & \degrade{$-$1.98} & 50.46 & \degrade{$-$3.58} & 64.31 & \imp{+0.34} \\
French & 0.14 & \imp{$-$0.15} & 70.68 & \imp{+3.16} & 57.51 & \imp{+0.00} & 50.83 & \degrade{$-$2.27} & 46.23 & \degrade{$-$3.14} & 58.04 & \imp{+0.29} \\
Italian & 0.14 & \imp{$-$0.14} & 70.31 & \imp{+3.04} & 57.33 & \degrade{$-$0.35} & 50.62 & \degrade{$-$1.90} & 45.18 & \degrade{$-$3.15} & 57.68 & \imp{+0.23} \\
Korean & 0.13 & \imp{$-$0.18} & 63.67 & \imp{+3.93} & 49.25 & \degrade{$-$0.23} & 41.65 & \degrade{$-$3.05} & 37.55 & \degrade{$-$3.93} & 49.87 & \imp{+0.25} \\
Portuguese & 0.14 & \imp{$-$0.16} & 70.30 & \imp{+3.39} & 59.29 & \imp{+0.45} & 52.44 & \degrade{$-$1.38} & 47.11 & \degrade{$-$2.39} & 58.95 & \imp{+0.87} \\
Russian & 0.13 & \imp{$-$0.16} & 66.81 & \imp{+3.74} & 54.81 & \imp{+0.20} & 48.14 & \degrade{$-$1.91} & 43.05 & \degrade{$-$3.34} & 54.90 & \imp{+0.61} \\
Spanish & 0.12 & \imp{$-$0.16} & 70.91 & \imp{+2.91} & 58.98 & \imp{+0.28} & 52.24 & \degrade{$-$1.51} & 47.19 & \degrade{$-$2.73} & 59.05 & \imp{+0.57} \\
\textbf{Avg.} & 0.13 & \imp{$-$0.16} & 68.41 & \imp{+3.39} & 56.36 & \imp{+0.18} & 49.51 & \degrade{$-$1.90} & 44.49 & \degrade{$-$3.13} & 56.42 & \imp{+0.46} \\
\cmidrule{1-13}
\multicolumn{13}{l}{\footnotesize\textit{\soft}} \\
Arabic & 0.21 & \imp{$-$0.18} & 53.84 & \imp{+1.74} & 46.95 & \imp{+0.92} & 41.77 & \degrade{$-$0.19} & 39.29 & \degrade{$-$0.20} & 46.19 & \imp{+1.12} \\
Chinese & 0.11 & \imp{$-$0.14} & 66.76 & \imp{+1.62} & 56.45 & \imp{+0.79} & 51.49 & \imp{+0.36} & 47.25 & \degrade{$-$0.58} & 56.90 & \imp{+0.36} \\
German & 0.10 & \imp{$-$0.15} & 67.58 & \imp{+0.53} & 58.79 & \degrade{$-$0.06} & 52.82 & \degrade{$-$0.72} & 49.03 & \degrade{$-$0.81} & 58.49 & \imp{+0.12} \\
English & 0.08 & \imp{$-$0.15} & 74.29 & \imp{+0.92} & 65.12 & \imp{+0.74} & 59.23 & \degrade{$-$0.24} & 53.35 & \degrade{$-$0.70} & 64.59 & \imp{+0.62} \\
French & 0.15 & \imp{$-$0.14} & 68.76 & \imp{+1.25} & 57.93 & \imp{+0.42} & 52.56 & \degrade{$-$0.54} & 48.83 & \degrade{$-$0.54} & 58.36 & \imp{+0.61} \\
Italian & 0.14 & \imp{$-$0.15} & 68.33 & \imp{+1.05} & 57.90 & \imp{+0.21} & 52.49 & \degrade{$-$0.03} & 47.55 & \degrade{$-$0.78} & 58.00 & \imp{+0.56} \\
Korean & 0.15 & \imp{$-$0.17} & 60.92 & \imp{+1.18} & 49.36 & \degrade{$-$0.12} & 43.54 & \degrade{$-$1.16} & 40.61 & \degrade{$-$0.87} & 49.89 & \imp{+0.28} \\
Portuguese & 0.15 & \imp{$-$0.15} & 68.32 & \imp{+1.41} & 59.43 & \imp{+0.59} & 53.76 & \degrade{$-$0.06} & 48.99 & \degrade{$-$0.51} & 58.94 & \imp{+0.87} \\
Russian & 0.12 & \imp{$-$0.17} & 64.21 & \imp{+1.14} & 55.11 & \imp{+0.51} & 49.76 & \degrade{$-$0.29} & 45.94 & \degrade{$-$0.45} & 55.00 & \imp{+0.71} \\
Spanish & 0.12 & \imp{$-$0.16} & 69.32 & \imp{+1.32} & 59.16 & \imp{+0.46} & 53.55 & \degrade{$-$0.20} & 49.21 & \degrade{$-$0.71} & 59.21 & \imp{+0.73} \\
\textbf{Avg.} & 0.13 & \imp{$-$0.16} & 66.23 & \imp{+1.22} & 56.62 & \imp{+0.45} & 51.10 & \degrade{$-$0.31} & 47.00 & \degrade{$-$0.62} & 56.56 & \imp{+0.60} \\
\cmidrule{1-13}
\multicolumn{13}{l}{\footnotesize\textit{\hard}} \\
Arabic & 0.22 & \imp{$-$0.16} & 53.77 & \imp{+1.67} & 47.16 & \imp{+1.13} & 41.63 & \degrade{$-$0.33} & 38.22 & \degrade{$-$1.28} & 45.97 & \imp{+0.90} \\
Chinese & 0.12 & \imp{$-$0.13} & 66.52 & \imp{+1.38} & 56.46 & \imp{+0.79} & 51.02 & \degrade{$-$0.11} & 46.59 & \degrade{$-$1.24} & 56.66 & \imp{+0.12} \\
German & 0.12 & \imp{$-$0.13} & 67.40 & \imp{+0.35} & 58.95 & \imp{+0.11} & 52.65 & \degrade{$-$0.89} & 47.91 & \degrade{$-$1.93} & 58.23 & \degrade{$-$0.13} \\
English & 0.09 & \imp{$-$0.13} & 74.02 & \imp{+0.65} & 65.22 & \imp{+0.84} & 58.90 & \degrade{$-$0.57} & 52.41 & \degrade{$-$1.63} & 64.28 & \imp{+0.31} \\
French & 0.16 & \imp{$-$0.13} & 68.90 & \imp{+1.38} & 58.34 & \imp{+0.83} & 52.42 & \degrade{$-$0.68} & 47.65 & \degrade{$-$1.72} & 58.27 & \imp{+0.52} \\
Italian & 0.16 & \imp{$-$0.13} & 68.32 & \imp{+1.04} & 58.26 & \imp{+0.58} & 52.29 & \degrade{$-$0.24} & 46.79 & \degrade{$-$1.55} & 57.89 & \imp{+0.45} \\
Korean & 0.15 & \imp{$-$0.16} & 60.88 & \imp{+1.14} & 49.68 & \imp{+0.20} & 43.22 & \degrade{$-$1.49} & 39.47 & \degrade{$-$2.01} & 49.69 & \imp{+0.07} \\
Portuguese & 0.17 & \imp{$-$0.14} & 67.43 & \imp{+0.52} & 59.01 & \imp{+0.18} & 53.20 & \degrade{$-$0.61} & 47.47 & \degrade{$-$2.04} & 58.17 & \imp{+0.10} \\
Russian & 0.14 & \imp{$-$0.15} & 64.03 & \imp{+0.96} & 55.38 & \imp{+0.77} & 49.44 & \degrade{$-$0.61} & 44.84 & \degrade{$-$1.55} & 54.71 & \imp{+0.42} \\
Spanish & 0.13 & \imp{$-$0.14} & 69.13 & \imp{+1.13} & 59.43 & \imp{+0.74} & 53.35 & \degrade{$-$0.40} & 48.35 & \degrade{$-$1.58} & 58.99 & \imp{+0.51} \\
\textbf{Avg.} & 0.15 & \imp{$-$0.14} & 66.04 & \imp{+1.02} & 56.79 & \imp{+0.62} & 50.81 & \degrade{$-$0.59} & 45.97 & \degrade{$-$1.65} & 56.29 & \imp{+0.33} \\
\cmidrule{1-13}
\multicolumn{13}{l}{\footnotesize\textit{Segment Embed Average}} \\
Arabic & 0.25 & \imp{$-$0.14} & 39.25 & \degrade{$-$12.85} & 37.19 & \degrade{$-$8.85} & 32.32 & \degrade{$-$9.64} & 30.13 & \degrade{$-$9.36} & 35.71 & \degrade{$-$9.36} \\
Chinese & 0.08 & \imp{$-$0.17} & 51.34 & \degrade{$-$13.80} & 46.22 & \degrade{$-$9.44} & 46.72 & \degrade{$-$4.41} & 45.19 & \degrade{$-$2.64} & 47.47 & \degrade{$-$9.07} \\
German & 0.25 & \degrade{+0.00} & 48.91 & \degrade{$-$18.14} & 47.70 & \degrade{$-$11.14} & 44.51 & \degrade{$-$9.03} & 41.09 & \degrade{$-$8.75} & 46.50 & \degrade{$-$11.86} \\
English & 0.25 & \degrade{+0.03} & 58.16 & \degrade{$-$15.21} & 54.24 & \degrade{$-$10.14} & 49.90 & \degrade{$-$9.57} & 46.08 & \degrade{$-$7.96} & 53.43 & \degrade{$-$10.53} \\
French & 0.26 & \imp{$-$0.03} & 51.28 & \degrade{$-$16.23} & 48.75 & \degrade{$-$8.76} & 44.97 & \degrade{$-$8.14} & 40.82 & \degrade{$-$8.55} & 47.58 & \degrade{$-$10.17} \\
Italian & 0.23 & \imp{$-$0.06} & 50.60 & \degrade{$-$16.68} & 47.99 & \degrade{$-$9.69} & 45.29 & \degrade{$-$7.24} & 41.25 & \degrade{$-$7.08} & 47.27 & \degrade{$-$10.18} \\
Korean & 0.28 & \imp{$-$0.03} & 43.10 & \degrade{$-$16.64} & 39.86 & \degrade{$-$9.62} & 36.45 & \degrade{$-$8.25} & 33.31 & \degrade{$-$8.17} & 39.23 & \degrade{$-$10.39} \\
Portuguese & 0.28 & \imp{$-$0.02} & 51.82 & \degrade{$-$15.09} & 48.50 & \degrade{$-$10.33} & 44.66 & \degrade{$-$9.16} & 41.18 & \degrade{$-$8.32} & 47.74 & \degrade{$-$10.33} \\
Russian & 0.28 & \imp{$-$0.01} & 46.39 & \degrade{$-$16.68} & 42.91 & \degrade{$-$11.69} & 39.37 & \degrade{$-$10.68} & 36.46 & \degrade{$-$9.93} & 42.40 & \degrade{$-$11.89} \\
Spanish & 0.26 & \imp{$-$0.01} & 52.25 & \degrade{$-$15.75} & 48.63 & \degrade{$-$10.07} & 44.86 & \degrade{$-$8.89} & 40.73 & \degrade{$-$9.19} & 47.88 & \degrade{$-$10.60} \\
\textbf{Avg.} & 0.24 & \imp{$-$0.04} & 49.31 & \degrade{$-$15.71} & 46.20 & \degrade{$-$9.97} & 42.90 & \degrade{$-$8.50} & 39.62 & \degrade{$-$8.00} & 45.52 & \degrade{$-$10.44} \\
\bottomrule
\end{tabular}
\end{adjustbox}
\caption{Multilingual retrieval with \textbf{bge-m3-dense}. Q1--Q4 denote nDCG@10 by query quartile; Overall is nDCG@10 across all queries. nDCG@10 values are $\times 100$. Deltas are relative to \textit{No Calibration}; \textcolor{teal}{teal} indicates improvement, \textcolor{red!70!black}{red} indicates degradation.}
\label{tab:appendix_multilingual_bgem3}
\end{table*}

\begin{table*}[t]
\centering
\setlength{\tabcolsep}{4pt}
\renewcommand{\arraystretch}{0.85}
\providecommand{\imp}[1]{{\scriptsize\color{teal}\,#1}}
\providecommand{\degrade}[1]{{\scriptsize\color{red!70!black}\,#1}}
\small
\begin{adjustbox}{max width=\textwidth}
\begin{tabular}{l
  r@{\hspace{1pt}}l
  r@{\hspace{1pt}}l
  r@{\hspace{1pt}}l
  r@{\hspace{1pt}}l
  r@{\hspace{1pt}}l
  r@{\hspace{1pt}}l}
\toprule
\textbf{Cross-lingual} & \multicolumn{2}{c}{} & \multicolumn{10}{c}{\textbf{nDCG@10}~$\uparrow$} \\
\cmidrule(lr){4-13}
\textbf{Retrieval} & \multicolumn{2}{c}{\textbf{PSI}~$\downarrow$} & \multicolumn{2}{c}{\textbf{Q1}} & \multicolumn{2}{c}{\textbf{Q2}} & \multicolumn{2}{c}{\textbf{Q3}} & \multicolumn{2}{c}{\textbf{Q4}} & \multicolumn{2}{c}{\textbf{Overall}~$\uparrow$} \\
\midrule[\heavyrulewidth]
\multicolumn{13}{l}{\footnotesize\textbf{gte-multilingual-base}\quad\textit{No Calibration}} \\
Arabic$\to$En & 0.28 &   & 41.11 &   & 31.88 &   & 24.59 &   & 21.80 &   & 31.28 &   \\
German$\to$En & 0.26 &   & 60.45 &   & 51.87 &   & 45.21 &   & 39.50 &   & 50.75 &   \\
French$\to$En & 0.24 &   & 63.98 &   & 54.34 &   & 47.21 &   & 41.98 &   & 53.47 &   \\
Italian$\to$En & 0.27 &   & 61.67 &   & 52.38 &   & 45.94 &   & 40.45 &   & 51.62 &   \\
Korean$\to$En & 0.23 &   & 44.02 &   & 33.13 &   & 25.56 &   & 22.27 &   & 32.88 &   \\
Portuguese$\to$En & 0.24 &   & 64.09 &   & 54.10 &   & 47.43 &   & 41.37 &   & 53.38 &   \\
Russian$\to$En & 0.28 &   & 55.56 &   & 45.32 &   & 37.18 &   & 32.99 &   & 44.45 &   \\
Spanish$\to$En & 0.24 &   & 64.84 &   & 54.80 &   & 47.94 &   & 42.19 &   & 54.06 &   \\
\textbf{Avg.} & 0.26 &   & 56.96 &   & 47.23 &   & 40.13 &   & 35.32 &   & 46.49 &   \\
\cmidrule{1-13}
\multicolumn{13}{l}{\footnotesize\textit{\uniform}} \\
Arabic$\to$En & 0.14 & \imp{$-$0.14} & 42.47 & \imp{+1.36} & 31.92 & \imp{+0.04} & 23.53 & \degrade{$-$1.06} & 19.76 & \degrade{$-$2.04} & 31.35 & \imp{+0.07} \\
German$\to$En & 0.16 & \imp{$-$0.10} & 61.76 & \imp{+1.31} & 52.07 & \imp{+0.20} & 44.43 & \degrade{$-$0.78} & 37.49 & \degrade{$-$2.01} & 50.91 & \imp{+0.16} \\
French$\to$En & 0.16 & \imp{$-$0.08} & 65.12 & \imp{+1.14} & 54.72 & \imp{+0.38} & 46.54 & \degrade{$-$0.67} & 39.90 & \degrade{$-$2.08} & 53.63 & \imp{+0.16} \\
Italian$\to$En & 0.18 & \imp{$-$0.09} & 62.46 & \imp{+0.79} & 52.73 & \imp{+0.35} & 45.04 & \degrade{$-$0.90} & 38.33 & \degrade{$-$2.12} & 51.61 & \degrade{$-$0.01} \\
Korean$\to$En & 0.13 & \imp{$-$0.10} & 44.93 & \imp{+0.91} & 32.98 & \degrade{$-$0.15} & 24.15 & \degrade{$-$1.41} & 19.64 & \degrade{$-$2.63} & 32.58 & \degrade{$-$0.30} \\
Portuguese$\to$En & 0.15 & \imp{$-$0.09} & 65.13 & \imp{+1.04} & 54.34 & \imp{+0.24} & 46.48 & \degrade{$-$0.95} & 39.01 & \degrade{$-$2.36} & 53.36 & \degrade{$-$0.02} \\
Russian$\to$En & 0.16 & \imp{$-$0.12} & 56.57 & \imp{+1.01} & 44.99 & \degrade{$-$0.33} & 35.92 & \degrade{$-$1.26} & 30.07 & \degrade{$-$2.92} & 44.13 & \degrade{$-$0.32} \\
Spanish$\to$En & 0.14 & \imp{$-$0.10} & 65.82 & \imp{+0.98} & 55.14 & \imp{+0.34} & 47.07 & \degrade{$-$0.87} & 40.00 & \degrade{$-$2.19} & 54.13 & \imp{+0.07} \\
\textbf{Avg.} & 0.15 & \imp{$-$0.11} & 58.03 & \imp{+1.07} & 47.36 & \imp{+0.13} & 39.15 & \degrade{$-$0.98} & 33.02 & \degrade{$-$2.30} & 46.46 & \degrade{$-$0.03} \\
\cmidrule{1-13}
\multicolumn{13}{l}{\footnotesize\textit{\soft}} \\
Arabic$\to$En & 0.14 & \imp{$-$0.14} & 41.13 & \imp{+0.02} & 33.04 & \imp{+1.16} & 25.78 & \imp{+1.19} & 22.90 & \imp{+1.10} & 32.28 & \imp{+1.00} \\
German$\to$En & 0.14 & \imp{$-$0.12} & 60.76 & \imp{+0.31} & 53.10 & \imp{+1.23} & 47.11 & \imp{+1.90} & 41.22 & \imp{+1.72} & 52.10 & \imp{+1.35} \\
French$\to$En & 0.13 & \imp{$-$0.11} & 64.44 & \imp{+0.46} & 55.89 & \imp{+1.55} & 49.32 & \imp{+2.11} & 43.61 & \imp{+1.63} & 55.00 & \imp{+1.53} \\
Italian$\to$En & 0.16 & \imp{$-$0.11} & 61.63 & \degrade{$-$0.04} & 53.78 & \imp{+1.40} & 47.76 & \imp{+1.82} & 42.19 & \imp{+1.74} & 52.91 & \imp{+1.29} \\
Korean$\to$En & 0.11 & \imp{$-$0.12} & 43.43 & \degrade{$-$0.59} & 33.98 & \imp{+0.85} & 26.70 & \imp{+1.14} & 23.09 & \imp{+0.82} & 33.55 & \imp{+0.67} \\
Portuguese$\to$En & 0.13 & \imp{$-$0.11} & 64.36 & \imp{+0.27} & 55.54 & \imp{+1.44} & 49.17 & \imp{+1.74} & 42.80 & \imp{+1.43} & 54.71 & \imp{+1.33} \\
Russian$\to$En & 0.15 & \imp{$-$0.13} & 55.48 & \degrade{$-$0.08} & 46.26 & \imp{+0.94} & 38.61 & \imp{+1.43} & 34.13 & \imp{+1.14} & 45.46 & \imp{+1.01} \\
Spanish$\to$En & 0.12 & \imp{$-$0.12} & 65.22 & \imp{+0.38} & 56.30 & \imp{+1.50} & 49.75 & \imp{+1.81} & 43.75 & \imp{+1.56} & 55.53 & \imp{+1.47} \\
\textbf{Avg.} & 0.14 & \imp{$-$0.12} & 57.05 & \imp{+0.09} & 48.49 & \imp{+1.26} & 41.77 & \imp{+1.64} & 36.71 & \imp{+1.39} & 47.69 & \imp{+1.20} \\
\cmidrule{1-13}
\multicolumn{13}{l}{\footnotesize\textit{\hard}} \\
Arabic$\to$En & 0.16 & \imp{$-$0.12} & 41.25 & \imp{+0.14} & 33.07 & \imp{+1.19} & 26.00 & \imp{+1.41} & 22.65 & \imp{+0.85} & 32.28 & \imp{+1.00} \\
German$\to$En & 0.17 & \imp{$-$0.09} & 60.42 & \degrade{$-$0.03} & 52.91 & \imp{+1.04} & 46.81 & \imp{+1.60} & 40.73 & \imp{+1.23} & 51.77 & \imp{+1.02} \\
French$\to$En & 0.18 & \imp{$-$0.06} & 63.86 & \degrade{$-$0.12} & 55.34 & \imp{+1.00} & 48.64 & \imp{+1.43} & 42.76 & \imp{+0.78} & 54.33 & \imp{+0.86} \\
Italian$\to$En & 0.19 & \imp{$-$0.08} & 61.34 & \degrade{$-$0.33} & 53.53 & \imp{+1.15} & 47.36 & \imp{+1.42} & 41.35 & \imp{+0.90} & 52.47 & \imp{+0.85} \\
Korean$\to$En & 0.15 & \imp{$-$0.08} & 43.67 & \degrade{$-$0.35} & 34.08 & \imp{+0.95} & 26.66 & \imp{+1.10} & 22.63 & \imp{+0.36} & 33.50 & \imp{+0.62} \\
Portuguese$\to$En & 0.16 & \imp{$-$0.08} & 63.95 & \degrade{$-$0.14} & 55.15 & \imp{+1.05} & 48.70 & \imp{+1.27} & 41.97 & \imp{+0.60} & 54.19 & \imp{+0.81} \\
Russian$\to$En & 0.17 & \imp{$-$0.11} & 55.34 & \degrade{$-$0.22} & 46.13 & \imp{+0.81} & 38.51 & \imp{+1.33} & 33.32 & \imp{+0.33} & 45.15 & \imp{+0.70} \\
Spanish$\to$En & 0.16 & \imp{$-$0.08} & 64.70 & \degrade{$-$0.14} & 55.76 & \imp{+0.96} & 49.25 & \imp{+1.31} & 42.88 & \imp{+0.69} & 54.89 & \imp{+0.83} \\
\textbf{Avg.} & 0.17 & \imp{$-$0.09} & 56.82 & \degrade{$-$0.14} & 48.25 & \imp{+1.02} & 41.49 & \imp{+1.36} & 36.04 & \imp{+0.72} & 47.32 & \imp{+0.83} \\
\cmidrule{1-13}
\multicolumn{13}{l}{\footnotesize\textit{Segment Embed Average}} \\
Arabic$\to$En & 0.21 & \imp{$-$0.07} & 19.54 & \degrade{$-$21.57} & 25.37 & \degrade{$-$6.51} & 25.35 & \imp{+0.76} & 26.21 & \imp{+4.41} & 23.75 & \degrade{$-$7.53} \\
German$\to$En & 0.20 & \imp{$-$0.06} & 39.82 & \degrade{$-$20.63} & 44.29 & \degrade{$-$7.58} & 42.91 & \degrade{$-$2.30} & 41.47 & \imp{+1.97} & 42.32 & \degrade{$-$8.43} \\
French$\to$En & 0.18 & \imp{$-$0.06} & 44.23 & \degrade{$-$19.75} & 47.00 & \degrade{$-$7.34} & 45.09 & \degrade{$-$2.12} & 43.36 & \imp{+1.38} & 45.29 & \degrade{$-$8.18} \\
Italian$\to$En & 0.17 & \imp{$-$0.10} & 41.23 & \degrade{$-$20.44} & 44.67 & \degrade{$-$7.71} & 43.40 & \degrade{$-$2.54} & 41.64 & \imp{+1.19} & 43.01 & \degrade{$-$8.61} \\
Korean$\to$En & 0.24 & \degrade{+0.01} & 21.11 & \degrade{$-$22.91} & 25.99 & \degrade{$-$7.14} & 27.07 & \imp{+1.51} & 26.94 & \imp{+4.67} & 25.02 & \degrade{$-$7.86} \\
Portuguese$\to$En & 0.16 & \imp{$-$0.08} & 43.74 & \degrade{$-$20.35} & 46.34 & \degrade{$-$7.76} & 44.99 & \degrade{$-$2.44} & 42.75 & \imp{+1.38} & 44.82 & \degrade{$-$8.56} \\
Russian$\to$En & 0.20 & \imp{$-$0.08} & 32.33 & \degrade{$-$23.23} & 37.49 & \degrade{$-$7.83} & 36.94 & \degrade{$-$0.24} & 35.93 & \imp{+2.94} & 35.65 & \degrade{$-$8.80} \\
Spanish$\to$En & 0.18 & \imp{$-$0.06} & 44.67 & \degrade{$-$20.17} & 47.51 & \degrade{$-$7.29} & 45.35 & \degrade{$-$2.59} & 43.43 & \imp{+1.24} & 45.71 & \degrade{$-$8.35} \\
\textbf{Avg.} & 0.19 & \imp{$-$0.06} & 35.84 & \degrade{$-$21.13} & 39.83 & \degrade{$-$7.39} & 38.89 & \degrade{$-$1.24} & 37.72 & \imp{+2.40} & 38.20 & \degrade{$-$8.29} \\
\bottomrule
\end{tabular}
\end{adjustbox}
\caption{Cross-lingual retrieval with \textbf{gte-multilingual-base}. Q1--Q4 denote nDCG@10 by query quartile; Overall is nDCG@10 across all queries. nDCG@10 values are $\times 100$. Deltas are relative to \textit{No Calibration}; \textcolor{teal}{teal} indicates improvement, \textcolor{red!70!black}{red} indicates degradation.}
\label{tab:appendix_crosslingual_gte}
\end{table*}

\begin{table*}[t]
\centering
\setlength{\tabcolsep}{4pt}
\renewcommand{\arraystretch}{0.85}
\providecommand{\imp}[1]{{\scriptsize\color{teal}\,#1}}
\providecommand{\degrade}[1]{{\scriptsize\color{red!70!black}\,#1}}
\small
\begin{adjustbox}{max width=\textwidth}
\begin{tabular}{l
  r@{\hspace{1pt}}l
  r@{\hspace{1pt}}l
  r@{\hspace{1pt}}l
  r@{\hspace{1pt}}l
  r@{\hspace{1pt}}l
  r@{\hspace{1pt}}l}
\toprule
\textbf{Cross-lingual} & \multicolumn{2}{c}{} & \multicolumn{10}{c}{\textbf{nDCG@10}~$\uparrow$} \\
\cmidrule(lr){4-13}
\textbf{Retrieval} & \multicolumn{2}{c}{\textbf{PSI}~$\downarrow$} & \multicolumn{2}{c}{\textbf{Q1}} & \multicolumn{2}{c}{\textbf{Q2}} & \multicolumn{2}{c}{\textbf{Q3}} & \multicolumn{2}{c}{\textbf{Q4}} & \multicolumn{2}{c}{\textbf{Overall}~$\uparrow$} \\
\midrule[\heavyrulewidth]
\multicolumn{13}{l}{\footnotesize\textbf{bge-m3-dense}\quad\textit{No Calibration}} \\
Arabic$\to$En & 0.35 &   & 46.56 &   & 38.91 &   & 34.93 &   & 32.58 &   & 38.85 &   \\
German$\to$En & 0.27 &   & 62.48 &   & 55.70 &   & 52.57 &   & 48.80 &   & 55.48 &   \\
French$\to$En & 0.27 &   & 63.95 &   & 56.00 &   & 52.88 &   & 48.70 &   & 56.12 &   \\
Italian$\to$En & 0.27 &   & 63.45 &   & 55.53 &   & 52.27 &   & 48.22 &   & 55.62 &   \\
Korean$\to$En & 0.39 &   & 50.13 &   & 41.67 &   & 37.86 &   & 35.61 &   & 41.92 &   \\
Portuguese$\to$En & 0.28 &   & 64.31 &   & 56.48 &   & 52.91 &   & 48.58 &   & 56.35 &   \\
Russian$\to$En & 0.33 &   & 56.82 &   & 48.43 &   & 45.19 &   & 41.70 &   & 48.72 &   \\
Spanish$\to$En & 0.27 &   & 64.63 &   & 56.76 &   & 53.53 &   & 49.14 &   & 56.77 &   \\
\textbf{Avg.} & 0.30 &   & 59.04 &   & 51.19 &   & 47.77 &   & 44.17 &   & 51.23 &   \\
\cmidrule{1-13}
\multicolumn{13}{l}{\footnotesize\textit{\uniform}} \\
Arabic$\to$En & 0.15 & \imp{$-$0.20} & 49.86 & \imp{+3.30} & 38.03 & \degrade{$-$0.88} & 31.71 & \degrade{$-$3.22} & 28.86 & \degrade{$-$3.72} & 38.63 & \degrade{$-$0.22} \\
German$\to$En & 0.13 & \imp{$-$0.15} & 65.89 & \imp{+3.41} & 55.81 & \imp{+0.10} & 50.56 & \degrade{$-$2.01} & 45.19 & \degrade{$-$3.62} & 55.84 & \imp{+0.36} \\
French$\to$En & 0.11 & \imp{$-$0.16} & 67.42 & \imp{+3.46} & 56.05 & \imp{+0.04} & 50.93 & \degrade{$-$1.95} & 45.10 & \degrade{$-$3.61} & 56.47 & \imp{+0.35} \\
Italian$\to$En & 0.12 & \imp{$-$0.15} & 66.82 & \imp{+3.38} & 55.28 & \degrade{$-$0.24} & 50.09 & \degrade{$-$2.18} & 44.59 & \degrade{$-$3.63} & 55.86 & \imp{+0.24} \\
Korean$\to$En & 0.17 & \imp{$-$0.21} & 54.30 & \imp{+4.17} & 40.99 & \degrade{$-$0.68} & 34.50 & \degrade{$-$3.36} & 31.47 & \degrade{$-$4.14} & 41.99 & \imp{+0.07} \\
Portuguese$\to$En & 0.11 & \imp{$-$0.18} & 67.67 & \imp{+3.37} & 56.46 & \degrade{$-$0.02} & 50.66 & \degrade{$-$2.25} & 44.60 & \degrade{$-$3.99} & 56.55 & \imp{+0.20} \\
Russian$\to$En & 0.15 & \imp{$-$0.18} & 60.50 & \imp{+3.67} & 47.81 & \degrade{$-$0.63} & 42.13 & \degrade{$-$3.06} & 37.20 & \degrade{$-$4.50} & 48.62 & \degrade{$-$0.09} \\
Spanish$\to$En & 0.10 & \imp{$-$0.17} & 68.07 & \imp{+3.44} & 56.56 & \degrade{$-$0.20} & 51.13 & \degrade{$-$2.40} & 45.60 & \degrade{$-$3.54} & 56.99 & \imp{+0.22} \\
\textbf{Avg.} & 0.13 & \imp{$-$0.18} & 62.57 & \imp{+3.52} & 50.87 & \degrade{$-$0.31} & 45.21 & \degrade{$-$2.55} & 40.32 & \degrade{$-$3.84} & 51.37 & \imp{+0.14} \\
\cmidrule{1-13}
\multicolumn{13}{l}{\footnotesize\textit{\soft}} \\
Arabic$\to$En & 0.15 & \imp{$-$0.20} & 45.63 & \degrade{$-$0.93} & 38.00 & \degrade{$-$0.91} & 33.59 & \degrade{$-$1.34} & 31.81 & \degrade{$-$0.77} & 38.16 & \degrade{$-$0.70} \\
German$\to$En & 0.12 & \imp{$-$0.15} & 62.75 & \imp{+0.27} & 55.77 & \imp{+0.07} & 52.00 & \degrade{$-$0.57} & 47.74 & \degrade{$-$1.07} & 55.57 & \imp{+0.09} \\
French$\to$En & 0.12 & \imp{$-$0.16} & 64.25 & \imp{+0.29} & 56.12 & \imp{+0.11} & 52.49 & \degrade{$-$0.40} & 47.68 & \degrade{$-$1.03} & 56.24 & \imp{+0.12} \\
Italian$\to$En & 0.11 & \imp{$-$0.16} & 63.92 & \imp{+0.47} & 55.60 & \imp{+0.07} & 51.69 & \degrade{$-$0.58} & 47.39 & \degrade{$-$0.83} & 55.85 & \imp{+0.23} \\
Korean$\to$En & 0.18 & \imp{$-$0.20} & 49.72 & \degrade{$-$0.41} & 40.89 & \degrade{$-$0.78} & 36.61 & \degrade{$-$1.24} & 34.68 & \degrade{$-$0.93} & 41.46 & \degrade{$-$0.47} \\
Portuguese$\to$En & 0.11 & \imp{$-$0.17} & 64.55 & \imp{+0.25} & 56.55 & \imp{+0.07} & 52.55 & \degrade{$-$0.36} & 47.56 & \degrade{$-$1.03} & 56.48 & \imp{+0.14} \\
Russian$\to$En & 0.14 & \imp{$-$0.19} & 56.60 & \degrade{$-$0.23} & 48.15 & \degrade{$-$0.29} & 44.16 & \degrade{$-$1.03} & 40.64 & \degrade{$-$1.06} & 48.45 & \degrade{$-$0.26} \\
Spanish$\to$En & 0.10 & \imp{$-$0.17} & 65.03 & \imp{+0.40} & 56.78 & \imp{+0.02} & 53.13 & \degrade{$-$0.41} & 48.32 & \degrade{$-$0.82} & 56.97 & \imp{+0.20} \\
\textbf{Avg.} & 0.13 & \imp{$-$0.18} & 59.06 & \imp{+0.01} & 50.98 & \degrade{$-$0.21} & 47.03 & \degrade{$-$0.74} & 43.23 & \degrade{$-$0.94} & 51.15 & \degrade{$-$0.08} \\
\cmidrule{1-13}
\multicolumn{13}{l}{\footnotesize\textit{\hard}} \\
Arabic$\to$En & 0.16 & \imp{$-$0.19} & 45.34 & \degrade{$-$1.22} & 38.46 & \degrade{$-$0.45} & 33.80 & \degrade{$-$1.14} & 31.13 & \degrade{$-$1.45} & 38.12 & \degrade{$-$0.73} \\
German$\to$En & 0.14 & \imp{$-$0.14} & 62.49 & \imp{+0.00} & 56.00 & \imp{+0.29} & 51.84 & \degrade{$-$0.72} & 47.04 & \degrade{$-$1.76} & 55.39 & \degrade{$-$0.10} \\
French$\to$En & 0.13 & \imp{$-$0.15} & 64.01 & \imp{+0.06} & 56.48 & \imp{+0.47} & 52.26 & \degrade{$-$0.62} & 46.88 & \degrade{$-$1.82} & 56.09 & \degrade{$-$0.03} \\
Italian$\to$En & 0.13 & \imp{$-$0.14} & 63.66 & \imp{+0.22} & 55.76 & \imp{+0.23} & 51.57 & \degrade{$-$0.70} & 46.47 & \degrade{$-$1.75} & 55.60 & \degrade{$-$0.02} \\
Korean$\to$En & 0.20 & \imp{$-$0.19} & 49.37 & \degrade{$-$0.76} & 41.38 & \degrade{$-$0.29} & 36.60 & \degrade{$-$1.25} & 34.03 & \degrade{$-$1.57} & 41.37 & \degrade{$-$0.55} \\
Portuguese$\to$En & 0.13 & \imp{$-$0.15} & 64.31 & \imp{+0.00} & 56.88 & \imp{+0.40} & 52.42 & \degrade{$-$0.49} & 46.86 & \degrade{$-$1.72} & 56.33 & \degrade{$-$0.01} \\
Russian$\to$En & 0.16 & \imp{$-$0.17} & 56.23 & \degrade{$-$0.60} & 48.48 & \imp{+0.05} & 44.04 & \degrade{$-$1.16} & 39.73 & \degrade{$-$1.96} & 48.24 & \degrade{$-$0.47} \\
Spanish$\to$En & 0.12 & \imp{$-$0.16} & 64.82 & \imp{+0.19} & 57.02 & \imp{+0.26} & 52.75 & \degrade{$-$0.78} & 47.36 & \degrade{$-$1.78} & 56.72 & \degrade{$-$0.05} \\
\textbf{Avg.} & 0.15 & \imp{$-$0.16} & 58.78 & \degrade{$-$0.26} & 51.31 & \imp{+0.12} & 46.91 & \degrade{$-$0.86} & 42.44 & \degrade{$-$1.73} & 50.98 & \degrade{$-$0.25} \\
\cmidrule{1-13}
\multicolumn{13}{l}{\footnotesize\textit{Segment Embed Average}} \\
Arabic$\to$En & 0.29 & \imp{$-$0.06} & 29.21 & \degrade{$-$17.35} & 30.75 & \degrade{$-$8.16} & 29.41 & \degrade{$-$5.52} & 28.50 & \degrade{$-$4.08} & 29.73 & \degrade{$-$9.12} \\
German$\to$En & 0.24 & \imp{$-$0.03} & 46.54 & \degrade{$-$15.94} & 46.22 & \degrade{$-$9.48} & 44.28 & \degrade{$-$8.29} & 41.24 & \degrade{$-$7.56} & 45.24 & \degrade{$-$10.24} \\
French$\to$En & 0.27 & \degrade{+0.00} & 47.00 & \degrade{$-$16.95} & 46.49 & \degrade{$-$9.51} & 44.53 & \degrade{$-$8.35} & 41.37 & \degrade{$-$7.33} & 45.57 & \degrade{$-$10.55} \\
Italian$\to$En & 0.25 & \imp{$-$0.02} & 47.11 & \degrade{$-$16.34} & 46.06 & \degrade{$-$9.47} & 44.24 & \degrade{$-$8.03} & 40.88 & \degrade{$-$7.34} & 45.31 & \degrade{$-$10.31} \\
Korean$\to$En & 0.29 & \imp{$-$0.10} & 31.93 & \degrade{$-$18.20} & 33.21 & \degrade{$-$8.46} & 31.79 & \degrade{$-$6.07} & 30.74 & \degrade{$-$4.87} & 32.26 & \degrade{$-$9.66} \\
Portuguese$\to$En & 0.25 & \imp{$-$0.03} & 47.89 & \degrade{$-$16.42} & 46.78 & \degrade{$-$9.70} & 44.38 & \degrade{$-$8.53} & 41.19 & \degrade{$-$7.39} & 45.88 & \degrade{$-$10.47} \\
Russian$\to$En & 0.28 & \imp{$-$0.05} & 39.15 & \degrade{$-$17.67} & 39.17 & \degrade{$-$9.26} & 38.18 & \degrade{$-$7.01} & 35.69 & \degrade{$-$6.01} & 38.52 & \degrade{$-$10.20} \\
Spanish$\to$En & 0.25 & \imp{$-$0.05} & 48.65 & \degrade{$-$15.98} & 47.08 & \degrade{$-$9.68} & 44.83 & \degrade{$-$8.70} & 41.21 & \degrade{$-$7.93} & 46.30 & \degrade{$-$10.47} \\
\textbf{Avg.} & 0.26 & \imp{$-$0.04} & 42.19 & \degrade{$-$16.86} & 41.97 & \degrade{$-$9.21} & 40.21 & \degrade{$-$7.56} & 37.60 & \degrade{$-$6.56} & 41.10 & \degrade{$-$10.13} \\
\bottomrule
\end{tabular}
\end{adjustbox}
\caption{Cross-lingual retrieval with \textbf{bge-m3-dense}. Q1--Q4 denote nDCG@10 by query quartile; Overall is nDCG@10 across all queries. nDCG@10 values are $\times 100$. Deltas are relative to \textit{No Calibration}; \textcolor{teal}{teal} indicates improvement, \textcolor{red!70!black}{red} indicates degradation.}
\label{tab:appendix_crosslingual_bgem3}
\end{table*}

\begin{table*}[t]
\centering
\setlength{\tabcolsep}{5pt}
\renewcommand{\arraystretch}{0.9}
\small
\resizebox{\textwidth}{!}{%
\begin{tabular}{l cccc cccc}
\toprule
\textbf{nDCG@10} & \multicolumn{4}{c}{\textbf{Original}} & \multicolumn{4}{c}{\textbf{Shuffled}} \\
\cmidrule(lr){2-5}\cmidrule(lr){6-9}
\textbf{Dataset} & No Calib & \uniform & \soft & \hard & No Calib & \uniform & \soft & \hard \\
\midrule
\multicolumn{9}{l}{\textbf{gte-multilingual-base}} \\
ArguAna & \textbf{58.92} & 56.60 & 54.29 & 56.28 & 56.61 & \textbf{56.65} & 56.52 & 55.55 \\
ClimateFEVER & 39.76 & 39.68 & 39.71 & \textbf{40.52} & 37.66 & 39.11 & 38.01 & \textbf{39.83} \\
DBPedia & \textbf{57.36} & 57.15 & 57.29 & 57.13 & 57.15 & 57.00 & \textbf{57.17} & 56.97 \\
FEVER & 93.20 & 94.22 & \textbf{95.08} & 94.22 & 93.58 & 93.68 & \textbf{93.80} & 93.68 \\
FiQA2018 & \textbf{59.65} & 58.92 & 58.87 & 58.23 & 54.90 & 53.51 & \textbf{56.15} & 53.78 \\
HotpotQA & \textbf{72.19} & 71.64 & 70.45 & 71.73 & 69.98 & \textbf{70.18} & 69.69 & 70.08 \\
MSMARCO & \textbf{64.13} & 63.19 & 63.99 & 63.19 & 61.78 & 61.78 & 61.93 & \textbf{62.18} \\
NFCorpus & 34.60 & 35.75 & 34.32 & \textbf{36.14} & 35.01 & 35.26 & 34.68 & \textbf{35.27} \\
NQ & 69.39 & 68.95 & \textbf{70.23} & 69.10 & 70.64 & 69.95 & \textbf{71.01} & 69.95 \\
QuoraRetrieval & \textbf{93.98} & \textbf{93.98} & 93.71 & 93.71 & \textbf{93.98} & \textbf{93.98} & 93.71 & \textbf{93.98} \\
SCIDOCS & 41.16 & 41.27 & \textbf{41.53} & 40.96 & \textbf{41.44} & 40.75 & 40.58 & 40.70 \\
SciFact & 75.97 & \textbf{77.36} & 76.06 & 77.32 & 74.11 & 74.82 & 74.72 & \textbf{75.23} \\
Touche2020 & 50.39 & 50.75 & \textbf{53.24} & 51.98 & 50.96 & 48.73 & \textbf{51.01} & 49.22 \\
\cmidrule(lr){1-9}
\textbf{Avg.} & \textbf{62.36} & 62.27 & 62.21 & 62.35 & 61.37 & 61.18 & \textbf{61.46} & 61.26 \\
\midrule
\multicolumn{9}{l}{\textbf{bge-m3-dense}} \\
ArguAna & 50.69 & 48.66 & \textbf{52.20} & 49.62 & 47.78 & 46.86 & \textbf{49.61} & 47.60 \\
ClimateFEVER & 36.88 & 34.45 & \textbf{37.11} & 36.27 & \textbf{37.18} & 33.45 & 37.08 & 34.46 \\
DBPedia & 60.47 & \textbf{60.57} & 60.40 & 60.51 & \textbf{60.94} & 60.89 & 60.81 & 60.77 \\
FEVER & \textbf{92.36} & 90.50 & 90.31 & 91.23 & \textbf{85.82} & 82.67 & 84.18 & 83.22 \\
FiQA2018 & 57.49 & 57.71 & 55.09 & \textbf{57.86} & 50.63 & 50.00 & 50.33 & \textbf{51.51} \\
HotpotQA & \textbf{81.88} & 80.92 & 80.85 & 81.21 & 81.13 & 81.65 & 81.20 & \textbf{82.06} \\
MSMARCO & 61.35 & 61.00 & \textbf{61.75} & 61.09 & 57.78 & 57.82 & 57.78 & \textbf{58.00} \\
NFCorpus & 33.10 & 33.35 & 33.20 & \textbf{33.66} & 32.01 & 32.30 & 31.96 & \textbf{32.31} \\
NQ & \textbf{68.49} & 66.02 & 67.51 & 66.30 & 67.27 & 67.22 & \textbf{68.60} & 66.88 \\
QuoraRetrieval & \textbf{95.37} & \textbf{95.37} & \textbf{95.37} & \textbf{95.37} & \textbf{95.51} & \textbf{95.51} & \textbf{95.51} & \textbf{95.51} \\
SCIDOCS & 36.19 & 35.80 & \textbf{36.45} & 35.91 & \textbf{35.48} & 32.40 & 33.55 & 33.39 \\
SciFact & 64.78 & \textbf{67.67} & 67.51 & 67.53 & 64.19 & \textbf{65.60} & 65.39 & 65.09 \\
Touche2020 & \textbf{48.66} & 45.54 & 46.35 & 47.46 & \textbf{47.55} & 44.96 & 45.07 & 47.00 \\
\cmidrule(lr){1-9}
\textbf{Avg.} & \textbf{60.59} & 59.87 & 60.32 & 60.31 & \textbf{58.71} & 57.81 & 58.54 & 58.29 \\
\bottomrule
\end{tabular}
}
\caption{Per-dataset nDCG@10 on NanoBEIR; \textbf{bold} marks the best method per dataset within each
condition. \textit{Shuffled} encodes each document with its sentences randomly
reshuffled before embedding. Values are $\times 100$.}
\label{tab:nanobeir_ndcg}
\end{table*}

\begin{table*}[t]
\centering
\resizebox{\textwidth}{!}{%
\setlength{\tabcolsep}{5pt}
\renewcommand{\arraystretch}{0.9}
\small
\begin{tabular}{l cccc cccc}
\toprule
\textbf{Recall@10} & \multicolumn{4}{c}{\textbf{Original}} & \multicolumn{4}{c}{\textbf{Shuffled}} \\
\cmidrule(lr){2-5}\cmidrule(lr){6-9}
\textbf{Dataset} & No Calib & \uniform & \soft & \hard & No Calib & \uniform & \soft & \hard \\
\midrule
\multicolumn{9}{l}{\textbf{gte-multilingual-base}} \\
ArguAna & \textbf{96.00} & \textbf{96.00} & 92.00 & \textbf{96.00} & 94.00 & \textbf{96.00} & 94.00 & \textbf{96.00} \\
ClimateFEVER & 45.13 & 44.30 & 45.40 & \textbf{46.30} & 44.80 & 45.80 & 45.80 & \textbf{46.80} \\
DBPedia & \textbf{32.50} & 32.29 & 32.25 & 32.29 & \textbf{33.22} & 33.17 & \textbf{33.22} & 33.17 \\
FEVER & \textbf{95.33} & \textbf{95.33} & \textbf{95.33} & \textbf{95.33} & \textbf{96.33} & 95.33 & 95.33 & 95.33 \\
FiQA2018 & 68.12 & \textbf{68.79} & 68.50 & 67.83 & 65.88 & 67.03 & \textbf{67.16} & 67.03 \\
HotpotQA & \textbf{72.00} & 71.00 & 70.00 & 71.00 & \textbf{72.00} & 71.00 & 71.00 & 71.00 \\
MSMARCO & \textbf{82.00} & \textbf{82.00} & \textbf{82.00} & \textbf{82.00} & \textbf{82.00} & \textbf{82.00} & \textbf{82.00} & \textbf{82.00} \\
NFCorpus & 16.45 & 16.67 & 16.13 & \textbf{16.76} & 13.75 & \textbf{14.16} & 13.58 & 14.14 \\
NQ & 85.00 & \textbf{87.00} & \textbf{87.00} & \textbf{87.00} & 86.00 & \textbf{87.00} & \textbf{87.00} & \textbf{87.00} \\
QuoraRetrieval & \textbf{100.00} & \textbf{100.00} & \textbf{100.00} & \textbf{100.00} & \textbf{100.00} & \textbf{100.00} & \textbf{100.00} & \textbf{100.00} \\
SCIDOCS & \textbf{42.17} & 41.77 & \textbf{42.17} & 41.37 & \textbf{40.57} & 39.87 & 40.27 & 40.27 \\
SciFact & 86.00 & \textbf{89.00} & 86.00 & \textbf{89.00} & \textbf{86.00} & \textbf{86.00} & \textbf{86.00} & \textbf{86.00} \\
Touche2020 & 30.34 & 30.63 & \textbf{32.10} & 31.25 & \textbf{31.52} & 29.94 & 31.13 & 30.37 \\
\cmidrule(lr){1-9}
\textbf{Avg.} & 65.47 & 65.75 & 65.30 & \textbf{65.86} & 65.08 & 65.18 & 65.12 & \textbf{65.32} \\
\midrule
\multicolumn{9}{l}{\textbf{bge-m3-dense}} \\
ArguAna & 84.00 & 78.00 & \textbf{86.00} & 82.00 & \textbf{86.00} & 82.00 & 84.00 & 84.00 \\
ClimateFEVER & 43.70 & 41.53 & \textbf{46.20} & 43.87 & 43.87 & 40.20 & \textbf{44.87} & 40.87 \\
DBPedia & \textbf{33.79} & 33.77 & 33.59 & 33.77 & \textbf{33.78} & 33.53 & 33.51 & 33.51 \\
FEVER & \textbf{97.33} & \textbf{97.33} & \textbf{97.33} & \textbf{97.33} & 94.67 & \textbf{95.33} & \textbf{95.33} & \textbf{95.33} \\
FiQA2018 & 62.90 & \textbf{65.30} & 62.74 & 64.64 & 62.07 & 61.35 & \textbf{62.35} & 60.69 \\
HotpotQA & \textbf{86.00} & 85.00 & 85.00 & 85.00 & 85.00 & \textbf{87.00} & 85.00 & \textbf{87.00} \\
MSMARCO & \textbf{86.00} & \textbf{86.00} & \textbf{86.00} & \textbf{86.00} & 80.00 & 80.00 & \textbf{82.00} & 80.00 \\
NFCorpus & 11.42 & 13.22 & 13.23 & \textbf{13.45} & \textbf{12.46} & 11.27 & 10.97 & 11.05 \\
NQ & \textbf{83.00} & 81.00 & \textbf{83.00} & \textbf{83.00} & 83.00 & 84.00 & \textbf{85.00} & 84.00 \\
QuoraRetrieval & \textbf{100.00} & \textbf{100.00} & \textbf{100.00} & \textbf{100.00} & \textbf{100.00} & \textbf{100.00} & \textbf{100.00} & \textbf{100.00} \\
SCIDOCS & \textbf{37.97} & 36.77 & 37.57 & 37.17 & \textbf{35.57} & 33.07 & 33.47 & 33.97 \\
SciFact & 79.00 & \textbf{81.00} & \textbf{81.00} & \textbf{81.00} & \textbf{77.00} & 75.00 & \textbf{77.00} & 75.00 \\
Touche2020 & \textbf{31.01} & 29.03 & 29.73 & 30.04 & \textbf{29.67} & 28.17 & 28.64 & 29.52 \\
\cmidrule(lr){1-9}
\textbf{Avg.} & 64.32 & 63.75 & \textbf{64.72} & 64.41 & \textbf{63.31} & 62.38 & 63.24 & 62.69 \\
\bottomrule
    \end{tabular}
}
\caption{Per-dataset Recall@10 on NanoBEIR; \textbf{bold} marks the best method per dataset within each
condition. \textit{Shuffled} encodes each document with its sentences randomly
reshuffled before embedding. Values are $\times 100$.}
\label{tab:nanobeir_recall}
\end{table*}

\end{document}